\documentclass[prd, letterpaper, amsmath, amssymb, preprintnumbers, showpacs, superscriptaddress, twocolumn, floatfix]{revtex4-1}
\usepackage{graphicx}
\usepackage{color}

\newcommand{\cA}{\ensuremath{\mathcal A} }
\newcommand{\cC}{\ensuremath{\mathcal C} }
\newcommand{\cD}{\ensuremath{\mathcal D} }
\newcommand{\cO}{\ensuremath{\mathcal O} }
\newcommand{\cZ}{\ensuremath{\mathcal Z} }
\newcommand{\Ga}{\ensuremath{\Gamma} }
\newcommand{\ga}{\ensuremath{\gamma} }
\newcommand{\si}{\ensuremath{\sigma} }
\newcommand{\fbar}{\ensuremath{\overline f} }
\newcommand{\psibar}{\ensuremath{\overline\psi} }
\newcommand{\chidof}{\ensuremath{\mbox{$\chi^2/\text{d.o.f.}$}}}

\newcommand{\mres}{\ensuremath{m_{\rm res}} }
\newcommand{\X}{\ensuremath{\!\times\!} }
\newcommand{\gsim}{\ensuremath{\gtrsim} }
\newcommand{\lsim}{\ensuremath{\lesssim} }
\newcommand{\Tr}[1]{\ensuremath{\mbox{Tr}\left[ #1 \right]} }
\newcommand{\vev}[1]{\ensuremath{\left\langle #1 \right\rangle} }
\newcommand{\pbp}{\ensuremath{\vev{\psibar\psi}} }
\newcommand{\ME}[3]{\ensuremath{\langle #1 \left| #2 \right| #3 \rangle} }
\newcommand{\eq}[1]{Eq.~(\ref{#1})}
\newcommand{\fig}[1]{Figure~\ref{#1}}
\newcommand{\tab}[1]{Table~\ref{#1}}
\newcommand{\secref}[1]{Section~\ref{#1}}
\newcommand{\refcite}[1]{Ref.~\cite{#1}}

\begin{document}
\title{Lattice simulations with eight flavors of domain wall fermions in SU(3) gauge theory}

\author{T.~Appelquist}
\affiliation{Department of Physics, Sloane Laboratory, Yale University, New Haven, Connecticut 06520, USA}
\author{R.~C.~Brower}
\affiliation{Department of Physics, Boston University, Boston, Massachusetts 02215, USA}
\author{G.~T.~Fleming}
\affiliation{Department of Physics, Sloane Laboratory, Yale University, New Haven, Connecticut 06520, USA}
\author{J.~Kiskis}
\affiliation{Department of Physics, University of California, Davis, California 95616, USA}
\author{M.~F.~Lin}
\affiliation{Argonne Leadership Computing Facility, Argonne National Laboratory, Argonne, Illinois 60439, USA}
\affiliation{Computational Science Center, Brookhaven National Laboratory, Upton, New York 11973, USA}
\author{E.~T.~Neil}
\affiliation{Department of Physics, University of Colorado, Boulder, Colorado 80309, USA}
\affiliation{RIKEN--BNL Research Center, Brookhaven National Laboratory, Upton, New York 11973, USA}
\author{J.~C.~Osborn}
\affiliation{Argonne Leadership Computing Facility, Argonne National Laboratory, Argonne, Illinois 60439, USA}
\author{C.~Rebbi}
\affiliation{Department of Physics, Boston University, Boston, Massachusetts 02215, USA}
\author{E.~Rinaldi}
\affiliation{Lawrence Livermore National Laboratory, Livermore, California 94550, USA}
\author{D.~Schaich}
\affiliation{Department of Physics, Syracuse University, Syracuse, New York 13244, USA}
\author{C.~Schroeder}
\affiliation{Lawrence Livermore National Laboratory, Livermore, California 94550, USA}
\author{S.~Syritsyn}
\affiliation{RIKEN--BNL Research Center, Brookhaven National Laboratory, Upton, New York 11973, USA}
\author{G.~Voronov}
\affiliation{Department of Physics, Sloane Laboratory, Yale University, New Haven, Connecticut 06520, USA}
\author{P.~Vranas}
\affiliation{Lawrence Livermore National Laboratory, Livermore, California 94550, USA}
\author{E.~Weinberg}
\affiliation{Department of Physics, Boston University, Boston, Massachusetts 02215, USA}
\author{O.~Witzel}
\affiliation{Center for Computational Science, Boston University, Boston, Massachusetts 02215, USA}
\collaboration{Lattice Strong Dynamics (LSD) Collaboration}
\noaffiliation

\date{24 October 2014}

\begin{abstract} 
  We study an SU(3) gauge theory with $N_f = 8$ degenerate flavors of light fermions in the fundamental representation.
  Using the domain wall fermion formulation, we investigate the light hadron spectrum, chiral condensate \pbp and electroweak $S$ parameter.
  We consider a range of light fermion masses on two lattice volumes at a single gauge coupling chosen so that IR scales approximately match those from our previous studies of the two- and six-flavor systems.
  Our results for the $N_f = 8$ spectrum suggest spontaneous chiral symmetry breaking, though fits to the fermion mass dependence of spectral quantities do not strongly disfavor the hypothesis of mass-deformed infrared conformality.
  Compared to $N_f = 2$ we observe a significant enhancement of \pbp relative to the symmetry breaking scale $F$, similar to the situation for $N_f = 6$.
  The reduction of the $S$ parameter, related to parity doubling in the vector and axial-vector channels, is also comparable to our six-flavor results.
\end{abstract}


\maketitle

\section{Introduction} 
The discovery of a Higgs particle at the Large Hadron Collider~\cite{Aad:2012tfa, Chatrchyan:2012ufa} was a major step towards the longstanding goal of determining the mechanism of electroweak symmetry breaking.
The properties of this particle are so far consistent with the predictions of the standard model~\cite{Chatrchyan:2013lba, Aad:2013wqa}, but could also result from new strong dynamics at or above the TeV scale.
Walking technicolor theories, in which approximately conformal dynamics produce a slowly-running gauge coupling and a large mass anomalous dimension across a wide range of energy scales~\cite{Holdom:1981rm, Yamawaki:1985zg, Appelquist:1986an}, are potential candidates to produce a light composite Higgs boson~\cite{Appelquist:2013sia}.
Numerical lattice gauge theory calculations are a crucial non-perturbative tool to study such strongly-interacting gauge theories from first principles.
In this paper we present results from lattice investigations of SU(3) gauge theory with $N_f = 8$ fundamental fermions, a candidate walking theory.

In SU($N$) gauge theories with $N_f$ massless fermions in the fundamental representation, chiral symmetry breaks spontaneously and the system confines if $N_f$ is sufficiently small.
When $N_f$ reaches a certain value $N_f^{(c)}$, with $N_f^{(c)} < N_f^{(AF)}$ at which asymptotic freedom is lost, the theory flows to a chirally symmetric conformal fixed point in the infrared (IRFP)~\cite{Caswell:1974gg, Banks:1981nn}.
The region $N_f^{(c)} \leq N_f < N_f^{(AF)}$ is called the conformal window for SU($N$) with fundamental fermions.
Around the upper end of the conformal window, $N_f \lsim N_f^{(AF)}$, the IRFP is weakly coupled and can be investigated perturbatively.
The fixed point moves to stronger coupling as $N_f$ decreases, motivating lattice studies of non-perturbative conformal or near-conformal dynamics for $N_f \sim N_f^{(c)}$.

Many lattice calculations have been performed to search for precise values of $N_f^{(c)}$, and more generally to explore the range of possible phenomena in these strongly-coupled gauge theories (cf.~the recent review~\cite{Kuti:2014LAT} and references therein).
For SU(3) gauge theories with fundamental fermions, these studies have focused on $N_f = 6$, 8, 10 and 12.
Although the 6-flavor theory exhibits interesting dynamical differences compared to QCD~\cite{Appelquist:2009ka, Appelquist:2010xv, Appelquist:2012sm, Miura:2012zqa}, there is little doubt that it is chirally broken.
Studies with larger $N_f$ are less conclusive.
Continuum estimates that $8 \lsim N_f^{(c)} \lsim 12$~\cite{Appelquist:1996dq, Appelquist:1999hr, Bashir:2013zha} make these difficult investigations particularly interesting.

For $N_f = 8$, several lattice studies~\cite{Appelquist:2007hu, Appelquist:2009ty, Deuzeman:2008sc, Fodor:2009wk, Hasenfratz:2010fi, Jin:2010vm} concluded that the theory most likely undergoes spontaneous chiral symmetry breaking.
More recently, Refs.~\cite{Cheng:2013eu, Cheng:2013bca} reported that the 8-flavor system possesses a large effective mass anomalous dimension across a wide range of energy scales.
The LatKMI Collaboration is investigating the light meson spectrum of the theory~\cite{Aoki:2013xza}, arguing that at lighter fermion masses $0.015 \leq m \leq 0.04$ the spectrum may be described by chiral perturbation theory, while data at heavier $0.05 \leq m \leq 0.16$ appear to exhibit some remnant of IR conformality despite chiral symmetry breaking.
They also find that the flavor-singlet scalar Higgs particle can be as light as the pseudoscalar meson (the would-be pion) throughout this range of $m$~\cite{Aoki:2014oha}.
Preliminary results from a large-scale USBSM project could not clearly confirm spontaneous chiral symmetry breaking with fermion masses as light as $m = 0.004$ on a $48^3\X96$ lattice volume~\cite{Schaich:2013eba}.

In contrast to the lattice studies summarized above, which all employ staggered fermions, we investigate $N_f = 8$ using the domain wall fermion formulation that possesses improved continuum-like chiral and flavor symmetries.
This work is the latest addition to our extensive investigations of SU(3) gauge theories with $N_f = 2$, 6, 8 and 10 flavors of degenerate domain wall fermions~\cite{Appelquist:2009ka, Appelquist:2010xv, Appelquist:2012sm, Appelquist:2012nz, Appelquist:2013ms}.
Domain wall fermions are more computationally expensive than staggered fermions, which limits the statistics we can obtain and is the reason we have not yet determined the fermion-line-disconnected contributions to flavor-singlet observables.
In the next section we summarize our 8-flavor simulations, which generate ensembles of gauge configurations for a range of light fermion masses on two lattice volumes, at a single gauge coupling chosen so that IR scales approximately match for all $N_f$.

We use these ensembles to investigate the light hadron spectrum, chiral condensate \pbp and electroweak $S$ parameter.
\secref{sec:spectrum} presents our spectrum analyses, first reviewing the determination of hadron masses and decay constants.
Steady growth in the ratio of the vector meson mass $M_V$ compared to the pseudoscalar mass $M_P$ as we approach the chiral limit suggests that chiral symmetry breaks spontaneously for $N_f = 8$.
Although we find the flavor non-singlet scalar meson to be heavy, $M_{a_0} > M_V$, we have not yet determined the mass of the more interesting flavor-singlet Higgs particle.
When we confront the fermion mass dependence of spectral quantities with expressions motivated by either spontaneous chiral symmetry breaking or mass-deformed IR conformality, we obtain comparable fit quality in each case.

In \secref{sec:pbp} we explore the enhancement of \pbp relative to the symmetry breaking scale $F$, which is of interest in the context of fermion mass generation.
We find a significant enhancement of the ratio $\pbp / F^3$ for $N_f = 8$ compared to $N_f = 2$, similar to results we previously reported for $N_f = 6$~\cite{Appelquist:2009ka, Fleming:2013tra}.
Finally, we study the electroweak $S$ parameter in \secref{sec:s}, also discussing the related issue of parity doubling in the vector ($V$) and axial-vector ($A$) channels.
We follow the approach of \refcite{Appelquist:2010xv} to calculate $S$ from the transverse $V$--$A$ vacuum polarization function.
At the range of masses we can access on our lattice ensembles, we observe parity doubling and a reduction in $S$ that are also comparable to six-flavor results from \refcite{Appelquist:2010xv}.
We summarize our conclusions and prospects for further progress in \secref{sec:conclusion}.
The Appendix provides additional information about thermalization, auto-correlations and the topological charge.

\section{\label{sec:sim}Simulation Details} 
\subsection{Parameters and algorithms} 
Our calculations are performed with the domain wall fermion (DWF) formulation~\cite{Shamir:1993zy, Furman:1994ky}, where an auxiliary fifth dimension separates the left-handed and right-handed chiralities.
We thereby obtain good chiral symmetry even at non-zero lattice spacing, with only a small chiral symmetry breaking effect quantified as the residual mass $\mres$.
With a non-zero input fermion mass $m_f$, the effective fermion mass is $m = m_f + \mres$ in the DWF formulation.

Using the Iwasaki gauge action~\cite{Iwasaki:2011np}, we tune the bare gauge coupling to $\beta \equiv 6 / g_0^2 = 1.95$ to obtain $M_{V0} \approx 0.2$ in lattice units, where $M_{V0} \equiv \lim_{m \to 0} M_V$ is the linear extrapolation of the vector meson mass to the chiral limit~\footnote{Such a linear chiral extrapolation of $M_V$ assumes that chiral symmetry is spontaneously broken, since in IR-conformal systems all masses vanish in the infinite-volume chiral limit.  The assumption of spontaneous chiral symmetry breaking appears reasonable for $N_f \leq 8$.  Because we tuned $\beta$ with a limited amount of initial data, our preliminary determination $M_{V0} \approx 0.2$ had large uncertainties, and is consistent with the somewhat smaller final result in \tab{tab:Nf}.}.
This value of $M_{V0}$ approximately matches those used in our 2- and 6-flavor investigations~\cite{Appelquist:2009ka} (cf.~\tab{tab:Nf}), and is equivalent to having a relatively large UV cutoff scale $a^{-1} \approx 5M_{V0}$.
If the theory is confined and chirally broken, then $M_{V0}$ is related to the confinement scale.
Having a large ratio $a^{-1} / M_{V0}$ helps to separate the IR physics from the UV physics, particularly in a theory where the gauge coupling may be running slowly.

We consider two lattice volumes, $L^3\X T = 32^3\X64$ and $16^3\X32$, with parameters summarized in \tab{tab:sim_pars}.
The length of the fifth dimension is fixed to $L_s = 16$ for both volumes.
To check for possible thermalization or poor sampling effects~\cite{Appelquist:2012nz}, we generate two independent $32^3\X64$ ensembles for each of the two lightest masses, $m_f = 0.01$ and 0.015, one starting from a random (disordered) gauge configuration, the other from an ordered configuration.
In all of our analyses we use a jackknife procedure with 50-trajectory blocks to reduce the effects of auto-correlations.
The Appendix provides additional information about auto-correlations and the thermalization cuts listed in \tab{tab:sim_pars}.
As discussed in the Appendix, 50-trajectory jackknife blocks may not remove all auto-correlation effects, which could cause our statistical uncertainties to be underestimated.
Since we do not have enough data to use larger blocks consistently, we increase our error estimates by 25\% to account for this potential underestimation.

\begin{table}[htbp]
  \begin{center}
    \begin{tabular}{cclccc}
      \hline
      $L^3\X T$   & $m_f$ & Start & Traj.   & Therm.  & Blocks  \\
      \hline
                  & 0.010 & dis   &  785    & 610     &  3      \\
                  & 0.010 & ord   & 2032    & 610     & 28      \\ 
                  & 0.015 & dis   & 1279    & 510     & 15      \\
      $32^3\X64$  & 0.015 & ord   & 1734    & 510     & 24      \\
                  & 0.020 & dis   & 1441    & 510     & 18      \\
                  & 0.025 & dis   & 1324    & 510     & 16      \\
                  & 0.030 & dis   & 1392    & 510     & 17      \\
      \hline
                  & 0.020 & ord   & 6665    & 610     & 31      \\
                  & 0.025 & ord   & 2780    & 610     & 19      \\
                  & 0.030 & ord   & 2460    & 610     & 13      \\
      $16^3\X32$  & 0.035 & ord   & 2860    & 610     & 29      \\
                  & 0.040 & ord   & 2400    & 610     & 29      \\
                  & 0.045 & ord   & 2219    & 610     & 22      \\
                  & 0.050 & ord   & 4219    & 610     & 14      \\ 
      \hline
    \end{tabular}
  \end{center}
  \caption{\label{tab:sim_pars}Simulation parameters, starting configuration and total number of trajectories in each of our DWF ensembles.  All ensembles use gauge coupling $\beta = 1.95$ and $L_s = 16$ in the fifth dimension.  All our analyses use a jackknife procedure with 50-trajectory blocks, so the number of blocks used for each ensemble depends on the corresponding thermalization cut.}
\end{table}

We generate gauge configurations using a Hybrid Monte Carlo algorithm with Hasenbuch mass preconditioning~\cite{Hasenbusch:2002ai}, in a style similar to \refcite{Aoki:2010dy}.
For each fermion determinant representing two degenerate flavors, an intermediate mass $m_I$ is used to precondition the input fermion mass $m_f$ and the Pauli--Villars mass $m_{PV} = 1$, resulting in a partition function of the form
\begin{equation}
  \label{eq:pf}
  \cZ[U] = \int [dU] \left\{\frac{\det\cD(m_f)}{\det\cD(m_I)}\frac{\det\cD(m_I)}{\det\cD(1)}\right\}^{N_f / 2} e^{-S_g[U]}
\end{equation}
for $N_f$ degenerate fermions.
Here $S_g[U]$ is the gauge action, $\cD(m) \equiv D^{\dag}(m) D(m)$ is the hermitian two-flavor domain wall Dirac operator~\cite{Blum:2000kn}, and $\cD(1)$ represents the DWF Pauli--Villars field.
We use intermediate mass $m_I = 0.1$ for all input $m_f$, and fix the trajectory length to be $\tau = 1$~molecular dynamics (MD) time unit.
While the intermediate mass term introduces additional pseudofermion fields, it helps to reduce the overall fermion force in the MD steps, making it possible to use larger step sizes while maintaining a good acceptance rate~\cite{Aoki:2004ht}.
Another reduction in computational cost comes from the use of the chronological inverter~\cite{Brower:1995vx}.
To avoid the loss of reversibility, we set stringent stopping conditions for the conjugate gradient matrix inversion: $10^{-9}$ for the MD evolution, and $10^{-10}$ for the Metropolis step.

\subsection{Residual mass and renormalization constants} 
\begin{figure}[htbp]
  \centering
  \includegraphics[width=\linewidth]{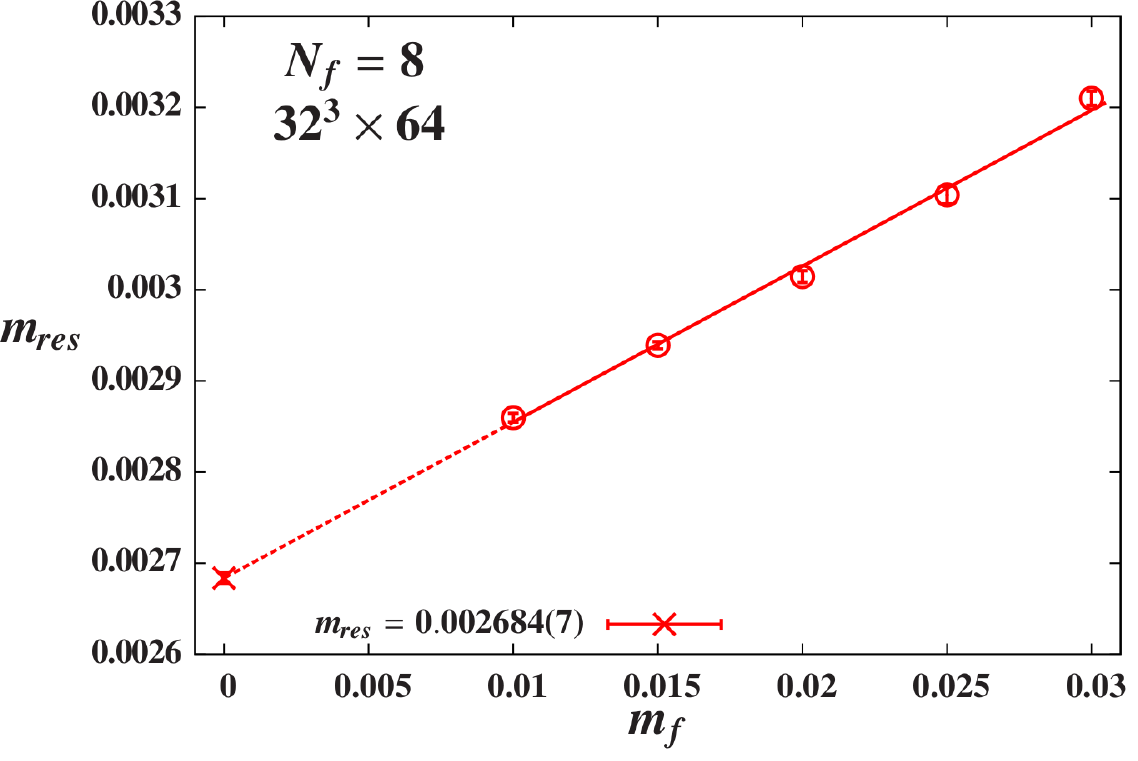}
  \caption{\label{fig:mres}Residual mass as a function of the input mass $m_f$ from our 8-flavor $32^3\X64$ ensembles, with linear $m_f \to 0$ extrapolation to determine $\mres = 0.002684(7)$.  The line is solid throughout the fit range $0.01 \leq m_f \leq 0.03$.}
\end{figure}

We calculate the residual mass following the standard procedure described in great detail by Refs.~\cite{Blum:2000kn, Allton:2008pn}.
The results from our $32^3\X64$ ensembles are recorded in \tab{tab:mres_ZA} and plotted in \fig{fig:mres}.
Defining \mres in the limit $m_f \to 0$, we obtain $\mres = 0.002684(7)$ from a simple linear extrapolation.

\begin{table}[htbp]
  \begin{center}
    \begin{tabular}{ccc}
      \hline
      $m_f$ & $\mres(m_f)\X10^3$  & $Z_A(m_f)$  \\
      \hline
      0.010 & 2.860(5)            & 0.70146(8)  \\
      0.015 & 2.939(4)            & 0.70173(5)  \\
      0.020 & 3.014(7)            & 0.70221(7)  \\
      0.025 & 3.104(10)           & 0.70273(5)  \\
      0.030 & 3.210(8)            & 0.70324(6)  \\
      \hline
    \end{tabular}
  \end{center}
  \caption{\label{tab:mres_ZA}Residual mass and axial-vector current renormalization constant as functions of the input mass $m_f$ from our 8-flavor $32^3\X64$ ensembles.}
\end{table}

\tab{tab:mres_ZA} also presents our $32^3\X64$ results for the DWF axial-vector current renormalization constant $Z_A$, determined from the procedure described in \refcite{Blum:2000kn}.
Our analyses below involve both this renormalization constant as well as the corresponding $Z_V$ for the vector current.
Because we define these renormalization constants in the chiral limit $m = m_f + \mres \to 0$, DWF obey the chiral symmetry relation $Z_A = Z_V$ up to lattice discretization effects of $\cO(a^2)$.
From a linear chiral extrapolation we find $Z_A = 0.70011(10)$, which we will use for both vector and axial-vector currents.

\tab{tab:Nf} summarizes our \mres and $Z_A$ results for all of $N_f = 2$, 6 and 8.
Screening effects from the additional fermions require that we work at stronger gauge couplings (smaller $\beta$) as $N_f$ increases, in order to maintain comparable IR scales such as $M_{V0}$ and the chirally-extrapolated baryon mass $M_{N0}$.
As a result, \mres increases by two orders of magnitude as we move from $N_f = 2$ to $N_f = 8$, while $Z_A$ moves farther from unity.
If $\beta$ is made too small, zero-temperature lattice calculations encounter a strong-coupling bulk transition at which the system becomes dominated by discretization artifacts.
In lattice QCD with $N_f = 2$ domain wall fermions and the Iwasaki gauge action we use, this occurs around $\beta \approx 1.95$.
As $N_f$ increases the bulk transition moves to slightly stronger coupling, $\beta \approx 1.75$ for $N_f = 8$.
Working with $\beta = 1.95$ safely on the weak-coupling side of this transition results in an 8-flavor $M_{V0}$ somewhat smaller than the values we obtained for $N_f = 2$ and 6.

\begin{table}[htbp]
  \begin{center}
    \begin{tabular}{cccccc}
      \hline
      $N_f$ & $\beta$ & $\mres\X10^3$ & $Z_A$         & $M_{V0}$      & $M_{N0}$      \\
      \hline
      2     & 2.70    &  0.0263(1)    &  0.85042(8)   &  0.2166(27)   &  0.2984(64)   \\
      6     & 2.10    &  0.8278(22)   &  0.72615(7)   &  0.1991(33)   &  0.2425(95)   \\
      8     & 1.95    & ~2.6836(72)~  & ~0.70011(10)~ & ~0.1710(36)~  &  0.2441(54)   \\
      \hline
    \end{tabular}
  \end{center}
  \caption{\label{tab:Nf}$N_f$ dependence of the residual mass $\mres$, the axial-vector current renormalization constant $Z_A$, and linear chiral extrapolations of the vector meson and baryon masses, $M_{V0}$ and $M_{N0}$, respectively.  Our results for $N_f = 2$ and 6 were first presented in \protect\refcite{Appelquist:2009ka}.}
\end{table}

We do not include the chirally-extrapolated pseudoscalar decay constant $F_{P0}$ among the IR scales that we attempt to match between systems with different $N_f$.
This is because $F_P$ is more sensitive to the form of the extrapolation, and needs to be analyzed using next-to-leading-order chiral perturbation theory (NLO $\chi$PT).
While we presented such an analysis for $N_f = 2$ in \refcite{Appelquist:2009ka}, we find that our 6- and 8-flavor data are not within the radius of convergence of NLO $\chi$PT.

\section{\label{sec:spectrum}Light Hadron Spectrum} 
\subsection{\label{sec:mesons}Meson masses and decay constants} 
Using the lattice ensembles discussed in the previous section, we analyze the light hadron spectrum, focusing especially on the pseudoscalar meson ($P$), the vector meson ($V$) and the axial-vector meson ($A$).
In chirally broken systems, the pseudoscalar meson is a pseudo-Nambu--Goldstone boson (PNGB), while the vector and axial-vector mesons may become more degenerate for theories near the conformal window.
In addition we consider the connected (flavor non-singlet) scalar meson ($a_0$), and in the next subsection we will study the lightest baryon ($N$).

We measure meson and baryon two-point correlators every 10 MD trajectories with both point ($p$) and Coulomb-gauge-fixed wall ($w$) sources, as well as $p$ and $w$ sinks.
Hence for each hadronic operator, we have four different types of two-point correlator, denoted as $C_{wp}$, $C_{ww}$, $C_{pw}$ and $C_{pp}$, where the subscripts indicate the sink and source, respectively.
To further increase statistics we also use two different source locations, $t_0 = 0$ and $t_0 = T / 2$, where $T$ is the temporal extent of the lattice.
In our analyses, we first average each correlator over the two source locations and block every 50 MD trajectories.
We then perform a simultaneous jackknife fit of the four averaged meson correlators to the form
\begin{align}
  C(t) & = \Tr{\vev{\sum_{\vec x} \psibar(\vec x, t) \Ga \tau^a \psi(\vec x, t) \psibar (\vec 0, 0) \Ga \tau^a \psi(\vec 0, 0)}}  \nonumber \\
       & = A \left[e^{-M t} + e^{-M(T - t)}\right],                                                                               \label{eq:expFit}
\end{align}
where $C(t)$ is projected to zero spatial momentum and the trace is over flavor.
(We normalize the flavor matrices $\tau^a$, $a = 1, 2, 3$ so that $\Tr{\tau^a \tau^b} = \frac{1}{2}\delta^{ab}$.)

For the pseudoscalar meson we can consider both $\Ga = \ga_5$ and $\Ga = \ga_4 \ga_5$, while $\Ga = \mathbb I$ for the $a_0$ scalar, $\Ga = \ga_i$ for the vector and $\Ga = \ga_i \ga_5$ for the axial-vector meson, with $i = 1$, 2, 3.
While each of the four source--sink combinations has an independent amplitude $A$, the meson mass $M$ is a common parameter in the simultaneous fit, in a way similar to \refcite{Allton:2008pn}.
Our limited statistics do not allow us to calculate a correlation matrix between the different source--sink combinations.
Because the pseudoscalar meson couples to both the $\psibar\ga_5 \psi$ and $\psibar\ga_4 \ga_5 \psi$ channels, our fit provides a common mass and eight amplitudes.
For the vector and axial-vector mesons, we first average over the three polarizations $i = 1$, 2, 3 to form a single correlator for each source--sink combination.
Our fits then provide a common mass and four amplitudes for each of the vector and axial-vector states.

\begin{figure*}[ht]
  \includegraphics[width=0.45\linewidth]{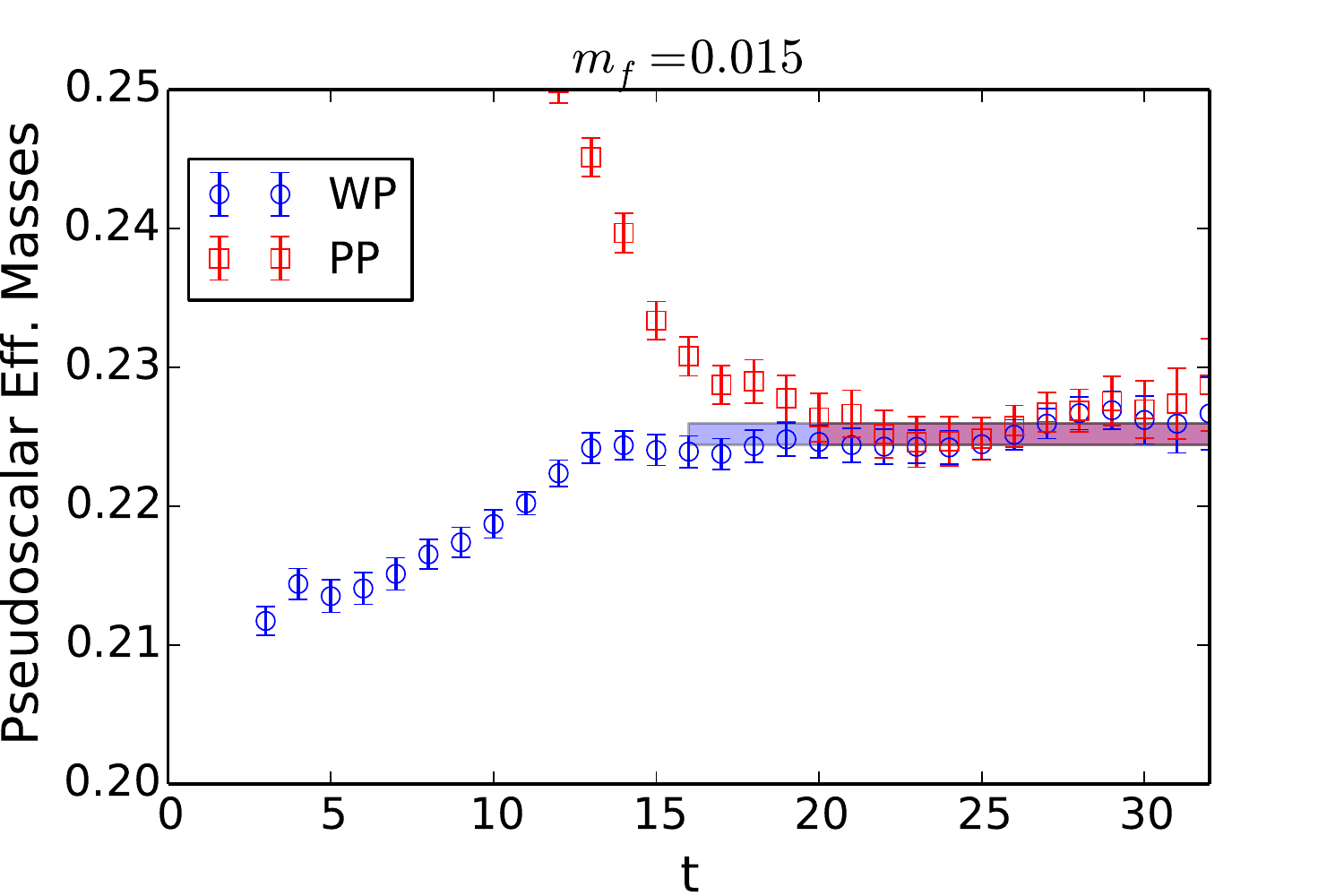}\hfill
  \includegraphics[width=0.45\linewidth]{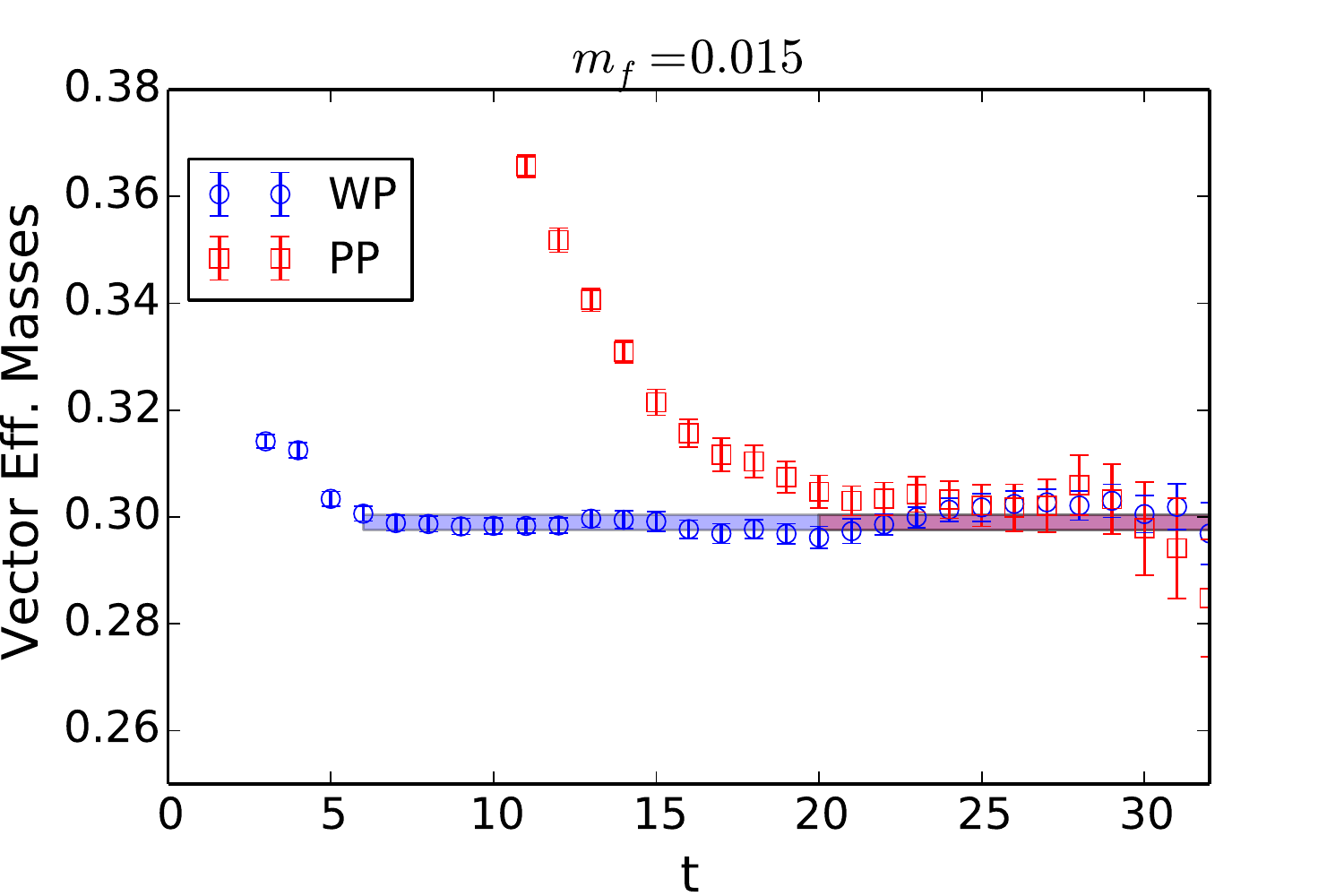}
  \caption{\label{fig:effm}Representative effective masses for the pseudoscalar meson (left) and vector meson (right), from our $32^3\X64$ ensembles with $m_f = 0.015$, combining ordered and disordered starts.  The shaded bands show the fit ranges and the uncertainties of the fit results. The red and blue regions in each band indicate the fit ranges for $C_{pp}(t)$ and $C_{wp}(t)$, respectively.}
\end{figure*}

Since different source--sink combinations have different excited-state contaminations, it is important to permit an independent fit range for each correlator in the simultaneous fit.
This allows us to make more efficient use of the available data while still avoiding excited-state effects.
To set fit ranges, we inspect the effective masses of the individual correlators to identify the onset of plateaus.
Since these bosonic correlators are symmetric around the middle timeslice $T / 2$ in the lattice, we ``fold'' the correlators by averaging $C(t)$ and $C(T - t)$ for $t \leq T / 2$.
Some representative effective mass plots from our $32^3\X64$ ensembles with $m_f = 0.015$ (combining ordered and disordered starts) are shown in \fig{fig:effm}.
To maintain some uniformity between different ensembles, we attempt to choose fairly conservative fit ranges that fall within plateaus for all $m_f$.
\tab{tab:fit_ranges} lists the resulting fit ranges used in our meson spectrum analyses.

\begin{table*}[htbp]
  \centering
  \begin{tabular}{cc|ccccc|c}
    \hline
                              &             &               &           & $32^3\X64$  &           &           & $16^3\X32$  \\
    \multicolumn{2}{c|}{Operator}           & $m_f = 0.010$ & 0.015     & 0.020       & 0.025     & 0.030     & All $m_f$   \\
    \hline
    $\psibar\ga_5 \psi$       & ~$C_{ww}$~  & [16, 32]      & [16, 32]  & [16, 32]    & [16, 32]  & [16, 32]  & [8, 16]     \\
    $\psibar\ga_5 \psi$       &  $C_{wp}$   & [16, 32]      & [16, 32]  & [16, 32]    & [16, 32]  & [16, 32]  & [8, 16]     \\
    $\psibar\ga_5 \psi$       &  $C_{pp}$   & [16, 32]      & [20, 32]  & [16, 32]    & [16, 32]  & [16, 32]  & [8, 16]     \\
    $\psibar\ga_5 \psi$       &  $C_{pw}$   & [16, 32]      & [16, 32]  & [16, 32]    & [16, 32]  & [16, 32]  & [8, 16]     \\
    $\psibar\ga_4 \ga_5 \psi$ &  $C_{ww}$   & [16, 32]      & [16, 32]  & [16, 32]    & [16, 32]  & [16, 32]  & [8, 16]     \\
    $\psibar\ga_4 \ga_5 \psi$ &  $C_{wp}$   & [16, 32]      & [16, 32]  & [16, 32]    & [16, 32]  & [16, 32]  & [8, 16]     \\
    $\psibar\ga_4 \ga_5 \psi$ &  $C_{pp}$   & [16, 32]      & [20, 32]  & [16, 32]    & [16, 32]  & [16, 32]  & [8, 16]     \\
    $\psibar\ga_4 \ga_5 \psi$ &  $C_{pw}$   & [16, 32]      & [16, 32]  & [16, 32]    & [16, 32]  & [16, 32]  & [8, 16]     \\
    \hline
    $\psibar\ga_i \psi$       &  $C_{ww}$   & [6, 32]       & [6, 32]   & [6, 32]     & [6, 22]   & [6, 22]   & [6, 16]     \\
    $\psibar\ga_i \psi$       &  $C_{wp}$   & [6, 32]       & [6, 32]   & [6, 32]     & [12, 32]  & [12, 32]  & [6, 16]     \\
    $\psibar\ga_i \psi$       &  $C_{pp}$   & [20, 32]      & [20, 32]  & [20, 32]    & [20, 32]  & [20, 32]  & [12, 16]    \\
    $\psibar\ga_i \psi$       &  $C_{pw}$   & [6, 32]       & [6, 32]   & [6, 32]     & [6, 22]   & [6, 22]   & [6, 16]     \\
    \hline
    $\psibar\ga_i \ga_5 \psi$ &  $C_{ww}$   & [10, 16]      & [5, 16]   & [10, 16]    & [5, 16]   & [5, 16]   & [10, 16]    \\
    $\psibar\ga_i \ga_5 \psi$ &  $C_{wp}$   & [10, 22]      & [5, 16]   & [10, 22]    & [5, 16]   & [5, 16]   & [10, 16]    \\
    $\psibar\ga_i \ga_5 \psi$ &  $C_{pp}$   & [16, 24]      & [20, 28]  & [16, 24]    & [20, 28]  & [20, 28]  & [12, 16]    \\
    $\psibar\ga_i \ga_5 \psi$ &  $C_{pw}$   & [6, 16]       & [6, 16]   & [6, 16]     & [6, 16]   & [6, 16]   & [6, 16]     \\
    \hline
    $\psibar \psi$            &  $C_{ww}$   & [15, 20]      & [6, 16]   & [15, 20]    & [13, 20]  & [13, 25]  & [10, 16]    \\
    $\psibar \psi$            &  $C_{wp}$   & [15, 20]      & [6, 16]   & [15, 20]    & [13, 20]  & [13, 25]  & [10, 16]    \\
    $\psibar \psi$            &  $C_{pp}$   & [15, 25]      & [6, 16]   & [15, 25]    & [20, 25]  & [20, 25]  & [10, 16]    \\
    $\psibar \psi$            &  $C_{pw}$   & [15, 20]      & [6, 16]   & [15, 20]    & [13, 20]  & [13, 20]  & [10, 16]    \\
    \hline
  \end{tabular}
  \caption{\label{tab:fit_ranges}Fit ranges in $t$ used to determine meson masses.}
\end{table*}

Turning to the flavor non-singlet decay constants, we define them through
\begin{align}
  \ME{0}{A^a_4}{\pi^a} & = -iZ_A \sqrt{2} F_P M_P \cr
  \ME{0}{V^a_i}{\rho^a} & = -iZ_V \sqrt{2} F_V M_V \epsilon_i \label{eq:fP} \\
  \ME{0}{A^a_i}{a_1^a} & = -iZ_A \sqrt{2} F_A M_A  \epsilon_i \nonumber
\end{align}
for the pseudoscalar, vector and axial-vector channels, respectively.
Here $\epsilon_i$ with $i = 1, 2, 3$ are polarization vectors, while $V_{\mu}^a(x) = \psibar(x) \ga_{\mu} \tau^a \psi(x)$ and $A_{\mu}^a(x) = \psibar(x) \ga_{\mu}\ga_5 \tau^a \psi(x)$ are the local (non-conserved) vector and axial-vector currents.
As discussed in the previous section, we use $Z_A = 0.70011(10)$ for both the vector and axial-vector current renormalization constants $Z_V$ and $Z_A$.
The above definitions are consistent with the conventions in \refcite{Peskin:1991sw}, with normalization such that the QCD pion decay constant is about 93~MeV.

The partially-conserved axial current (PCAC) relation also allows us to determine the pseudoscalar decay constant from the pseudoscalar matrix element~\cite{Blum:2000kn}.
For DWF, the PCAC relation is
\begin{equation}
  \partial_{\mu} \cA^a_{\mu}(x) = 2 (m_f + \mres) P^a(x),
\end{equation}
where $\cA_{\mu}^a(x)$ is the (partially-)conserved axial-vector current and $P^a(x) = \psibar(x) \ga_5 \tau^a \psi(x)$ is the local pseudoscalar current.
This allows us to replace the axial-vector matrix element in \eq{eq:fP} by the pseudoscalar matrix element, giving
\begin{equation}
  2 (m_f + \mres)\ME{0}{P^a}{\pi^a} = -i\sqrt{2} F_P M_P^2.
\end{equation}
The amplitudes we obtain from the simultaneous fits allow us to determine the decay constants in several different ways.
For each ensemble, we take the jackknife average of these different determinations as our final result.

As discussed in \secref{sec:sim}, for $m_f = 0.01$ and 0.015 we generate separate $32^3\X64$ ensembles using either ordered or disordered starting configurations, to check for possible bias from inadequate thermalization or from poor sampling of the topological sectors.
\tab{tab:start-comp} compares results for the meson masses and decay constants determined separately on these ordered- and disordered-start ensembles.
For both $m_f = 0.01$ and 0.015, the separate results agree well enough that we can combine the two Markov chains to perform our final analysis.
The final fit results for the meson masses and decay constants are shown in \tab{tab:fit_results_32c} for the $32^3\X64$ ensembles, and in \tab{tab:fit_results_16c} for the $16^3\X32$ ensembles.

\begin{table*}[htbp]
  \centering
  \begin{tabular}{cccccccc}
    \hline
    $m_f$   & Start         & $M_P$         & $M_V$         & $M_A$         & $F_P$         & $F_V$         & $F_A$         \\
    \hline
     0.010  &  dis          &  0.1983(66)   &  0.2786(66)   &  0.3441(52)   &   0.0285(7)   &  0.0438(6)    &  0.0445(32)   \\
     0.010  &  ord          &  0.1844(21)   &  0.2659(20)   &  0.3403(65)   &   0.0299(8)   &  0.0456(9)    &  0.0422(15)   \\
     \hline
     0.015  &  dis          &  0.2252(19)   &  0.3029(30)   &  0.4138(47)   &   0.0360(11)  &  0.0486(14)   &  0.0459(22)   \\
    ~0.015~ & ~ord~         & ~0.2251(13)~  & ~0.2964(20)~  & ~0.3915(36)~  &  ~0.0365(10)~ & ~0.0466(8)~   & ~0.0400(16)~  \\
    \hline
  \end{tabular}
  \caption{\label{tab:start-comp}Comparison of meson masses and decay constants from $32^3\X64$ ensembles with different starting configurations.}
\end{table*}

The reasonable agreement between $N_f = 8$ results from ordered and disordered starts is in stark contrast to the 10-flavor case~\cite{Appelquist:2012nz}, where we observed significant disagreements that we attributed to frozen topological charges $Q$.
While the topological charges sampled by our 8-flavor ensembles do not produce the desired gaussian distributions, $Q$ tunnels frequently, as we show in the Appendix.
This tunneling appears sufficient to eliminate the systematic discrepancy introduced when the topological charge is completely frozen.

\begin{table*}[htbp]
  \centering
  \begin{tabular}{cccccccc}
    \hline
    $m_f$   & $M_P$         & $M_V$         & $M_A$         & $M_{a_0}$    & $F_P$         & $F_V$         & $F_A$         \\
    \hline
     0.010  &  0.1853(21)   &  0.2664(21)   &  0.3411(64)   &  0.296(12)  &  0.0297(7)    &  0.0453(7)    &  0.0425(15)   \\
     0.015  &  0.2252(10)   &  0.2990(18)   &  0.3999(37)   &  0.381(8)   &  0.0364(7)    &  0.0474(7)    &  0.0421(12)   \\
     0.020  &  0.2582(14)   &  0.3363(27)   &  0.4414(102)  &  0.471(34)  &  0.0395(7)    &  0.0495(9)    &  0.0425(21)   \\
     0.025  &  0.2949(11)   &  0.3759(38)   &  0.5052(49)   &  0.474(17)  &  0.0456(9)    &  0.0541(14)   &  0.0427(32)   \\
    ~0.030~ & ~0.3280(12)~  & ~0.4131(35)~  & ~0.5502(38)~  & ~0.518(26)~ & ~0.0504(13)~  & ~0.0586(16)~  & ~0.0493(42)~  \\
    \hline
  \end{tabular}
  \caption{\label{tab:fit_results_32c}Results for meson masses and decay constants from $32^3\X64$ ensembles.}
\end{table*}

\begin{table*}[htbp]
  \centering
  \begin{tabular}{cccccccc}
    \hline
    $m_f$   & $M_P$         & $M_V$       & $M_A$       & $M_{a_0}$   & $F_P$         & $F_V$         & $F_A$       \\
    \hline
     0.020  &  0.3725(83)   &  0.468(13)  &  0.483(18)  &  0.383(6)   &  0.0305(6)    &  0.0722(29)   &  0.065(4)   \\
     0.025  &  0.3880(83)   &  0.485(4)   &  0.517(9)   &  0.423(8)   &  0.0378(12)   &  0.0745(16)   &  0.068(2)   \\
     0.030  &  0.3957(81)   &  0.486(13)  &  0.511(31)  &  0.439(23)  &  0.0459(14)   &  0.0775(29)   &  0.060(6)   \\
     0.035  &  0.3930(67)   &  0.501(6)   &  0.596(18)  &  0.497(15)  &  0.0536(15)   &  0.0779(17)   &  0.066(5)   \\
     0.040  &  0.4120(41)   &  0.524(7)   &  0.621(27)  &  0.559(31)  &  0.0590(13)   &  0.0801(16)   &  0.061(6)   \\
     0.045  &  0.4373(31)   &  0.550(6)   &  0.753(39)  &  0.678(65)  &  0.0641(12)   &  0.0852(22)   &  0.087(11)  \\
    ~0.050~ & ~0.4544(52)~  & ~0.557(5)~  & ~0.716(95)~ & ~0.609(50)~ & ~0.0661(12)~  & ~0.0829(27)~  & ~0.068(29)~ \\
    \hline
  \end{tabular}
  \caption{\label{tab:fit_results_16c}Results for meson masses and decay constants from $16^3\X32$ ensembles.}
\end{table*}

\subsection{Baryon mass} 
The zero-momentum projected two-point baryon correlator is
\begin{equation}
  B(t) = \sum_{\vec x} \vev{N(\vec x, t) N^\dagger (\vec 0, 0)},
\end{equation}
where the interpolating operator $N(x)$ is
\begin{equation}
  N(x) = \epsilon_{ijk} \left[ \psi_i^T(x) \cC \gamma_5 \psi_j(x) \right] \psi_k(x).
\end{equation}
In the above equation $i, j, k = 1, 2, 3$ are color indices and \cC is the charge conjugation operator.
The fermion fields $\psi_a(x)$ have anti-periodic boundary conditions in the time direction.
If we define the parity projection operators $P_{\pm} = (1 \pm \gamma_4)/2$, then on a lattice with temporal extent $T$, the large-$t$ behavior of $B(t)$ can be written as
\begin{equation}
  \begin{split}
    P_+ B(t) & = A_N e^{-M_N t} + A_{N^*} e^{-M_{N^*} (T - t)} \\
    P_- B(t) & = -A_{N^*} e^{-M_{N^*} t} - A_N e^{-M_N(T - t)},
  \end{split}
\end{equation}
where $N$ and $N^*$ represent the ground states of the baryon and of its parity partner.

For each two-point correlator measurement, we average the positive- and negative-parity-projected correlators to define
\begin{equation}
  B_N(t) = [P_+ B(t) - P_- B(T - t)] / 2.
\end{equation}
When $t \ll T$, the averaged correlator takes the simple exponential form
\begin{equation}
  B_N(t) \approx A_N e^{-M_N t},
\end{equation}
and we use single-exponential fits to determine the baryon mass $M_N$.
By considering $(T - t) \ll T$, one would obtain the mass of the baryon's parity partner $N^*$.
However, we could not reliably determine $M_{N^*}$ from our current data, and only report results for $M_N$.
We also present results only from our $32^3\X64$ ensembles, since the temporal extent of the $16^3\X32$ lattices is not large enough to provide reliable plateaus.

\begin{figure}[htbp]
  \centering
  \includegraphics[width=\linewidth]{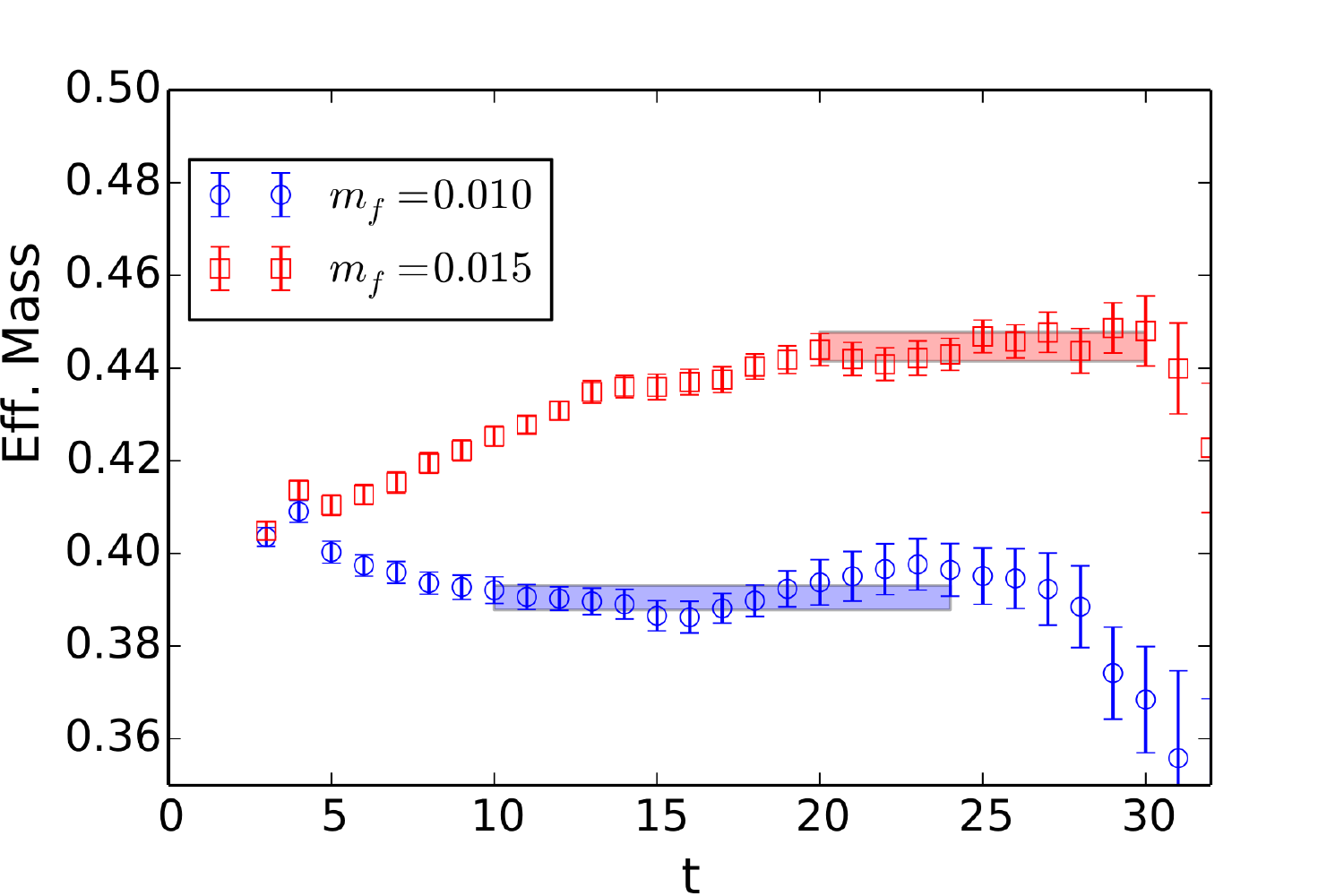}
  \caption{Representative baryon effective masses from our $32^3\X64$ ensembles with $m_f = 0.01$ and 0.015, combining ordered and disordered starts while considering only wall sources and point sinks.  The shaded bands indicate the fit ranges and the uncertainties of the fit results.}
  \label{fig:effm_N}
\end{figure}

We consider only wall sources with point sinks to determine the baryon mass, as this source--sink combination provides the best signal-to-noise ratio.
As in the analysis for the mesons, we average over the two source locations and block every 50 MD trajectories.
\fig{fig:effm_N} presents two representative sets of effective mass results (for our $32^3\X64$ ensembles with $m_f = 0.01$ and 0.015, combining ordered and disordered starts), which show that the approach to a plateau varies significantly for different $m_f$.
We choose fit ranges in $t$ by requiring that the fit results do not change beyond statistical uncertainties upon dropping the first or last points.
These fit ranges, and the corresponding baryon mass results, are tabulated in \tab{tab:nucmass}.
As for our meson spectrum results, we combine ordered- and disordered-start ensembles to determine the final results for $m_f = 0.01$ and 0.015.
We find good agreement between $M_N$ computed separately on the ordered- and disordered-start ensembles.

\begin{table}[htbp]
  \begin{tabular}{ccccc}
    \hline
    $m_f$   & Fit range             & $M_N$         & dis         & ord         \\
    \hline
    ~0.010~ & ~$10 \leq t \leq 24$~ & ~0.3905(33)~  & ~0.401(10)~ & ~0.390(3)~  \\
     0.015  &  $20 \leq t \leq 30$  &  0.4446(39)   &  0.446(6)   &  0.444(5)   \\
     0.020  &  $15 \leq t \leq 24$  &  0.4962(41)   &  ---        &  ---        \\
     0.025  &  $16 \leq t \leq 30$  &  0.5590(50)   &  ---        &  ---        \\
     0.030  &  $16 \leq t \leq 30$  &  0.6189(41)   &  ---        &  ---        \\
    \hline
  \end{tabular}
  \caption{\label{tab:nucmass}Fit ranges and results for the baryon mass from $32^3\X64$ ensembles.}
\end{table}

\begin{figure}[htbp]
  \centering
  \includegraphics[width=\linewidth]{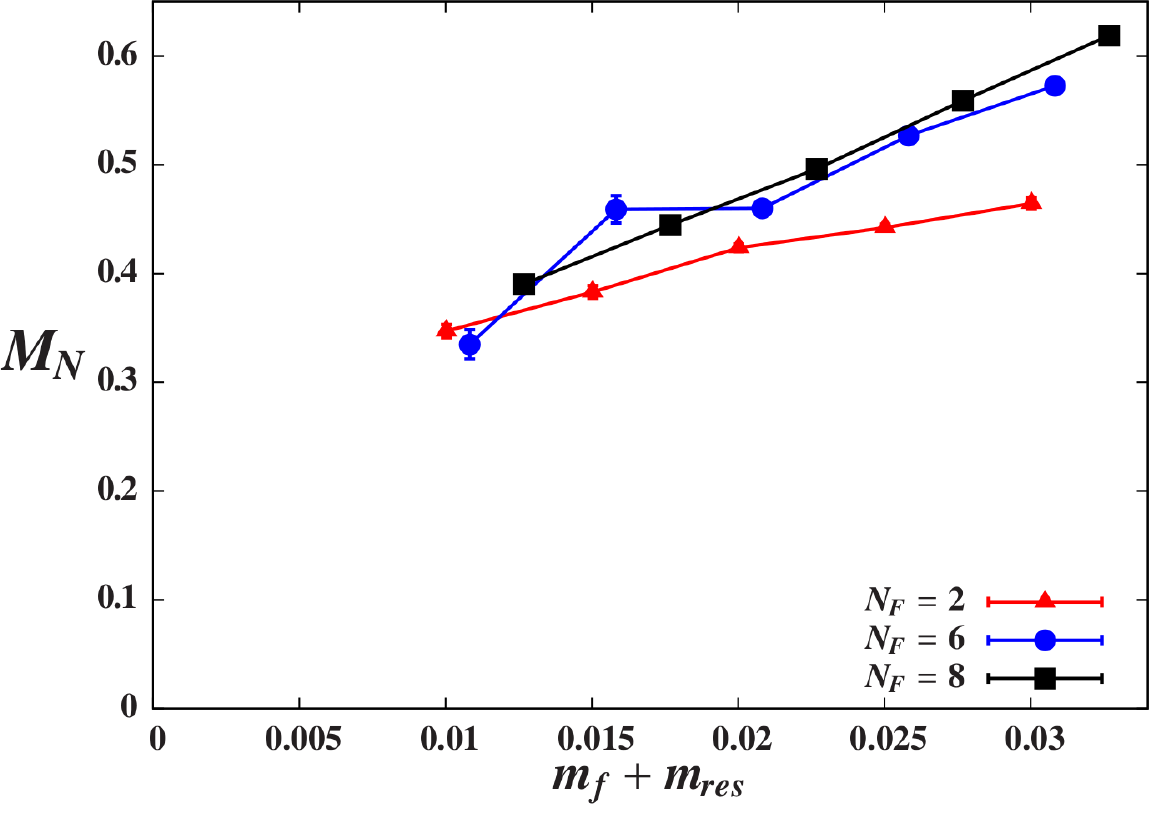}
  \caption{Baryon mass $M_N$ plotted against $m = m_f + \mres$ for $N_f = 2$, 6 and 8, each with $0.01 \leq m_f \leq 0.03$ on $32^3\X64$ lattices.  Lines connect points to guide the eye.}
  \label{fig:MN}
\end{figure}

In \fig{fig:MN} we plot our baryon mass results as functions of $m = m_f + \mres$ for all of our investigations with $N_f = 2$, 6 and 8.
Via the Feynman--Hellmann theorem, the $m$-dependence of $M_N$ is related to the baryon $\si_B$ term~\cite{WalkerLoud:2008bp, Appelquist:2014dja}
\begin{equation}
  M_N \sum_{\psi} f_{\psi}^{(B)} = \si_B = \vev{B|m\psibar\psi|B} = m\frac{\partial M_N}{\partial m}.
\end{equation}
Estimating $\frac{\partial M_N}{\partial m}$ as the slope of $M_N$ vs.\ $m$, we find that the 6- and 8-flavor slopes agree within uncertainties, and both are twice as large as the result for $N_f = 2$. 
In terms of the dimensionless $\frac{\si_B}{M_N} = \frac{m}{M_N}\frac{\partial M_N}{\partial m}$, we find $0.16 \lsim \frac{\si_B}{M_N} \lsim 0.37$ for $N_f = 2$, $0.35 \lsim \frac{\si_B}{M_N} \lsim 0.58$ for $N_f = 6$ and $0.37 \lsim \frac{\si_B}{M_N} \lsim 0.60$ for $N_f = 8$, with $0.01 \leq m_f \leq 0.03$ in each case.

\subsection{Chiral symmetry breaking} 
\begin{figure}[htbp]
  \centering
  \includegraphics[width=\linewidth]{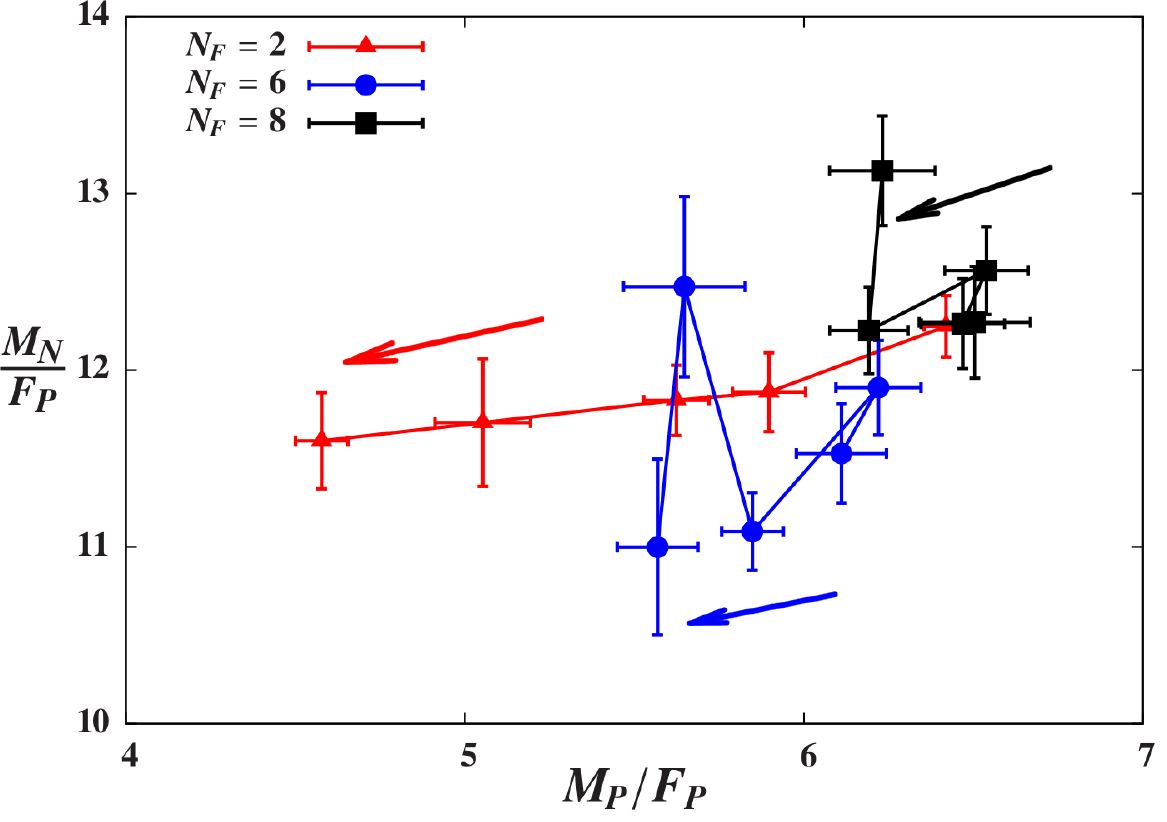}
  \caption{\label{fig:Edin}Edinburgh-style plot of $M_N / F_P$ vs.\ $M_P / F_P$, for $N_f = 2$, 6 and 8, each with $0.01 \leq m_f \leq 0.03$ on $32^3\X64$ lattices.  Lines connect points to guide the eye, and arrows indicate the direction of decreasing masses.  Our results for $N_f = 2$ and 6 were first presented in \protect\refcite{Appelquist:2009ka}.}
\end{figure}

We begin our discussion of the spectrum results by considering the Edinburgh-style plot in \fig{fig:Edin}, which presents the ratios $M_N / F_P$ vs.\ $M_P / F_P$ for all of our investigations with $N_f = 2$, 6 and 8.
For each $N_f$ we analyze $32^3\X64$ lattice ensembles with the same five input fermion masses $0.01 \leq m_f \leq 0.03$.
These correspond to $4.4 \lsim M_P L \lsim 7.8$ for $N_f = 2$, $5.4 \lsim M_P L \lsim 9.7$ for $N_f = 6$ and $5.9 \lsim M_P L \lsim 10.5$ for $N_f = 8$.
The ratios in \fig{fig:Edin} are designed to exaggerate finite-volume effects, which we expect to increase the masses while decreasing $F_P$, pushing the points up and to the right.
Our results do not show this behavior for any $N_f$.
The 2-flavor ratios move steadily to the left, as we would expect from spontaneous chiral symmetry breaking: the pseudoscalar becomes a massless NGB in the chiral limit, while $F_P$ and $M_N$ remain non-zero.
Although the 6-flavor results also move to the left, they do not move as much as do those for $N_f = 2$, and the 8-flavor points cluster in a small region of the plot.
The lightest point for $N_f = 8$ may hint at the onset of finite-volume effects, but such effects are not yet significant in these ratios.
In addition to providing evidence that finite-volume effects are under control, \fig{fig:Edin} illustrates some of the differences between the three systems with $N_f = 2$, 6 and 8.

\begin{figure*}[htbp]
  \includegraphics[width=0.45\linewidth]{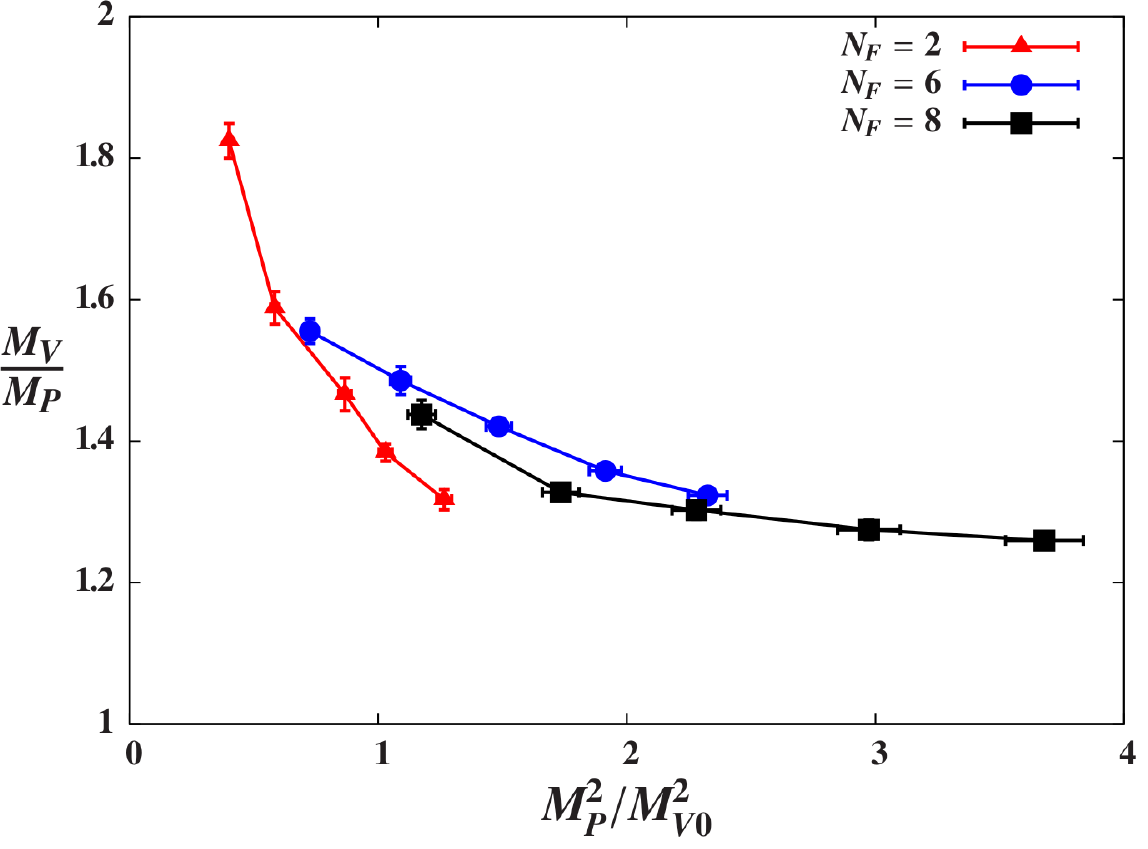}\hfill
  \includegraphics[width=0.45\linewidth]{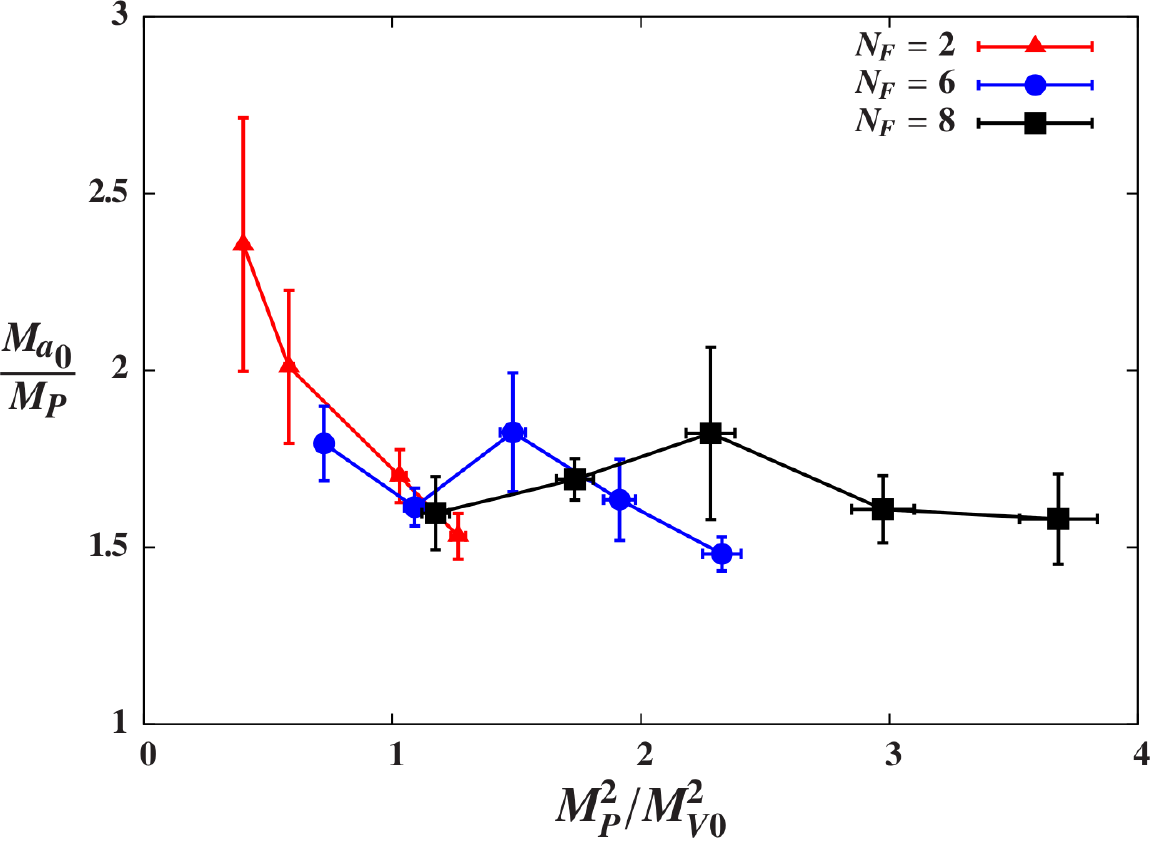}
  \caption{\label{fig:M_ov_MP}$M_V / M_P$ (left) and $M_{a_0} / M_P$ (right) plotted against $M_P^2 / M_{V0}^2$ for $N_f = 2$, 6 and 8, each with $0.01 \leq m_f \leq 0.03$ on $32^3\X64$ lattices.  Lines connect points to guide the eye.  Our $M_P$ and $M_V$ results for $N_f = 2$ and 6 were first presented in \protect\refcite{Appelquist:2009ka}.}
\end{figure*}

The ratios $M_V / M_P$ and $M_{a_0} / M_P$ provide similar illustrations, which we consider in \fig{fig:M_ov_MP}.
These quantities diverge in the chiral limit for chirally broken systems in which $M_P \to 0$ while $M_V$ and $M_{a_0}$ remain non-zero.
\fig{fig:M_ov_MP} again compares $N_f = 8$ with our earlier results for $N_f = 2$ and 6, considering the same $0.01 \leq m_f \leq 0.03$ on $32^3\X64$ lattices but now plotting vs.\ $M_P^2 / M_{V0}^2$.
The 2-flavor results for $M_V / M_P$ increase rapidly, as we expect for a chirally broken theory with non-zero fermion mass.
While this ratio does not grow so dramatically for $N_f = 6$ and 8, both show similar monotonic increases as $M_P$ decreases, suggesting spontaneous chiral symmetry breaking.
The 2-flavor results for the ratio $M_{a_0} / M_P$ are qualitatively similar, though the uncertainties are significantly larger.
(We could not reliably determine $M_{a_0}$ for $N_f = 2$ with $m_f = 0.02$, for which $M_P^2 / M_{V0}^2 \approx 0.87$.)
The 6- and 8-flavor results for $M_{a_0} / M_P$ are roughly constant within uncertainties.
This behavior may be related to parity doubling, an issue we will discuss further in the context of the electroweak $S$ parameter in \secref{sec:parity}.

Although we find the $a_0$ mass to be larger than the TeV-scale $M_V$ in the absence of finite-volume effects, this has no direct implications for the mass of the flavor-singlet scalar Higgs particle.
The flavor-singlet state is sensitive to fermion-line-disconnected contributions that are extremely expensive to compute, especially in the DWF formulation.
These disconnected contributions appear crucial to the LatKMI Collaboration's observation of a flavor-singlet scalar roughly degenerate with the pseudoscalar for $M_P \gsim 0.18$~\cite{Aoki:2014oha}.
As an alternative approach, we have explored gluonic operators that also couple to the scalar channel.
Our preliminary results suggest that our ensembles do not possess sufficient statistics to permit robust glueball analyses, another consequence of working with expensive DWF.
Because it is so important to determine whether the observed 125~GeV Higgs is consistent with new strong dynamics, in the future we plan to explore disconnected DWF calculations, to judge whether the available computational resources would provide reliable results.

\begin{figure*}[htbp]
  \includegraphics[width=0.45\linewidth]{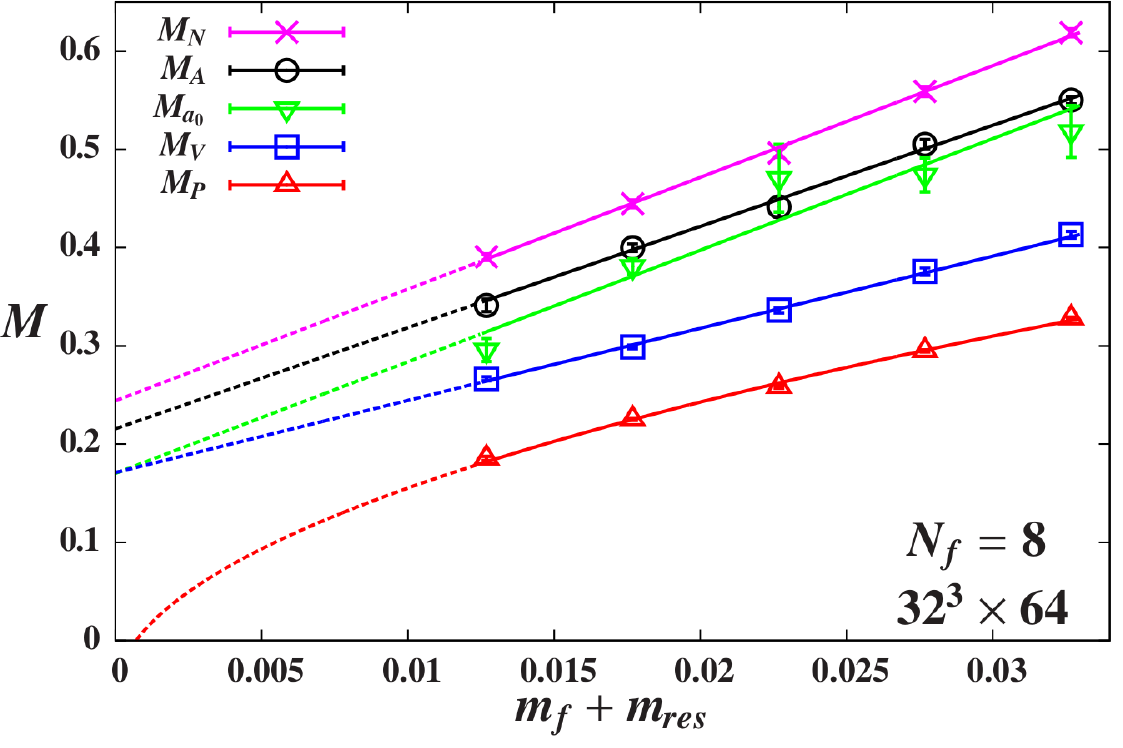}\hfill
  \includegraphics[width=0.45\linewidth]{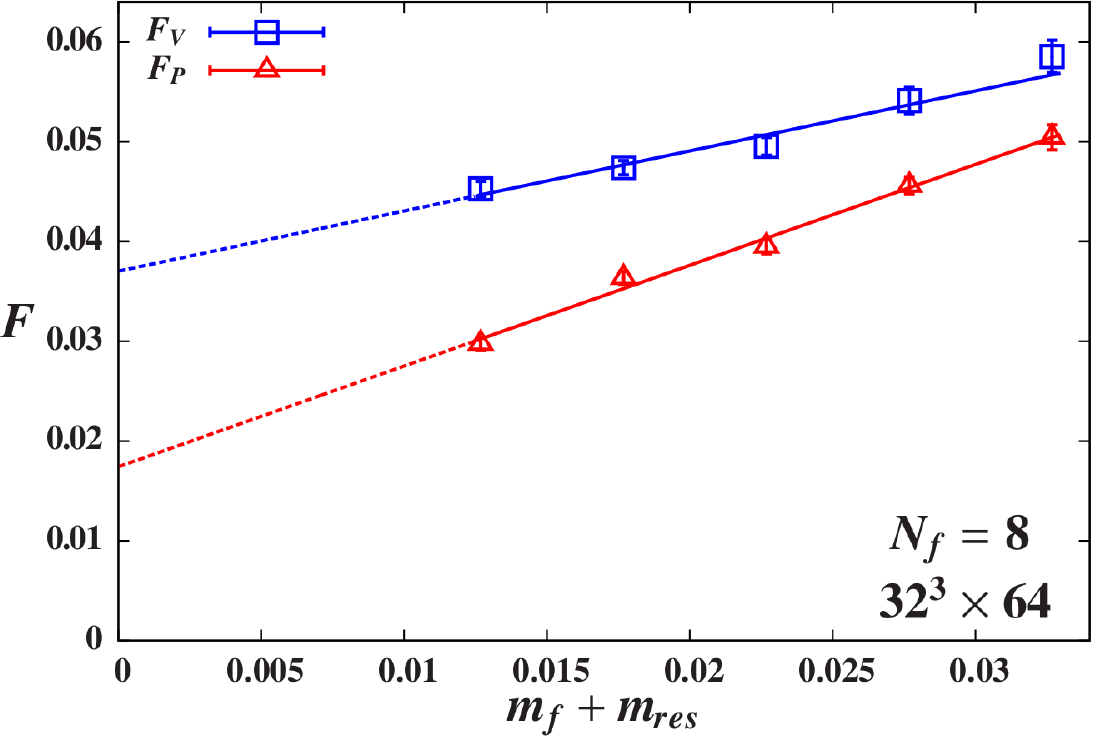}
  \caption{\label{fig:linFits}Fits of our $32^3\X64$ results for hadron masses (left) and decay constants (right) to the forms in \protect\eq{eq:linFits} motivated by spontaneous chiral symmetry breaking.  The lines are solid throughout the fit range $0.01 \leq m_f \leq 0.03$.}
\end{figure*}

If chiral symmetry does break spontaneously in the 8-flavor system, as \fig{fig:M_ov_MP} suggests, then leading-order $\chi$PT predicts that at small $m$ the PNGB mass squared is $M_P^2 \propto m$.
The other masses and decay constants should be well modeled by a constant plus a term linear in $m$.
We are not working at light enough masses to expect to resolve chiral logarithms.
We therefore further explore the hypothesis of spontaneous chiral symmetry breaking by considering fits to the following simple forms:
\begin{align}
  M_P^2 & = C_0^{(P)} + C_1^{(P)} m \cr
  M_{V, A, a_0, N} & = C_0^{(V, A, a_0, N)} + C_1^{(V, A, a_0, N)} m \label{eq:linFits} \\
  F_{P, V, A} & = D_0^{(P, V, A)} + D_1^{(P, V, A)} m. \nonumber
\end{align}
These fits are shown in \fig{fig:linFits}.
Their quality varies significantly depending on the observable considered, as indicated by the values $1 \lsim \chidof \lsim 10$ in the second column of \tab{tab:linFits}.

The third, fourth and fifth columns of \tab{tab:linFits} explore the fit-range dependence of these chiral extrapolations, by omitting the lightest point $m_f = 0.01$, the heaviest point $m_f = 0.03$, and both of these points, respectively.
Most quantities show relatively little sensitivity to the fit range, with the different intercepts agreeing within statistical uncertainties.
There is also no significant systematic trend in the \chidof~values as data points are omitted from the fits.
The significantly non-zero intercepts (and large \chidof) we find for the pseudoscalar mass squared indicate that our results are not well described by leading-order $\chi$PT, $M_P^2 \propto m$.
This behavior is consistent with our expectation that our 8-flavor data are not within the radius of convergence of $\chi$PT, which is also the case for $N_f = 6$ in this range of $m_f$~\cite{Appelquist:2009ka, Neil:2010sc}.

\begin{table*}[htbp]
  \begin{center}
    \begin{tabular}{c|cc|cc|cc|cc}
      \hline
                    & \multicolumn{2}{|c|}{~$0.01 \leq m_f \leq 0.03$~} & \multicolumn{2}{|c|}{~$0.01 \leq m_f \leq 0.025$~}  & \multicolumn{2}{|c|}{~$0.015 \leq m_f \leq 0.03$~}  & \multicolumn{2}{|c}{~$0.015 \leq m_f \leq 0.025$~}  \\
      ~Observable~  & \multicolumn{2}{|c|}{$\text{d.o.f.} = 3$}         & \multicolumn{2}{|c|}{$\text{d.o.f.} = 2$}           & \multicolumn{2}{|c|}{$\text{d.o.f.} = 2$}           & \multicolumn{2}{|c}{$\text{d.o.f.} = 1$}            \\
      \hline
      $M_P^2$       & $-$0.014(1)  & 4.8                                & $-$0.011(2)  & 3.1                                  & $-$0.016(2)  & 4.5                                  & $-$0.013(2)   & 4.3                                 \\
      $M_V$         &    0.171(4)  & 1.0                                &    0.174(5)  & 1.2                                  &    0.164(5)  & 0.1                                  &    0.164(8)   & 0.1                                 \\
      $M_A$         &    0.216(7)  & 0.9                                &    0.208(10) & 0.8                                  &    0.221(9)  & 0.8                                  &    0.214(14)  & 1.1                                 \\
      $M_{a_0}$     &    0.170(21) & 2.3                                &    0.155(25) & 2.8                                  &    0.216(29) & 0.8                                  &    0.210(37)  & 1.5                                 \\
      $M_N$         &    0.244(5)  & 1.0                                &    0.250(7)  & 0.7                                  &    0.235(9)  & 0.8                                  &    0.243(14)  & 1.2                                 \\
      $F_P$         &    0.017(1)  & 1.6                                &    0.017(1)  & 2.4                                  &    0.019(2)  & 1.6                                  &    0.020(2)   & 2.8                                 \\
      $F_V$         &    0.037(1)  & 1.3                                &    0.038(2)  & 0.8                                  &    0.035(2)  & 0.9                                  &    0.036(3)   & 1.1                                 \\
      $F_A$         &    0.040(3)  & 0.6                                &    0.042(4)  & 0.04                                 &    0.037(5)  & 0.6                                  &    0.041(6)   & 0.002                               \\
      \hline
    \end{tabular}
  \end{center}
  \caption{\label{tab:linFits}Intercepts and \chidof~for linear chiral extrapolations of the spectrum data in Tables~\protect\ref{tab:fit_results_32c} and \protect\ref{tab:nucmass}, using the forms in \protect\eq{eq:linFits} motivated by spontaneous chiral symmetry breaking and considering several different fit ranges.  The large \chidof~and non-zero intercepts for the pseudoscalar mass squared indicate that our results are not well described by leading-order $\chi$PT, $M_P^2 \propto m$.}
\end{table*}

\subsection{Conformal hypothesis} 
Even though \fig{fig:M_ov_MP} suggests that the 8-flavor system exhibits spontaneous chiral symmetry breaking, it is still worthwhile to consider the alternate hypothesis that the theory is IR conformal in the infinite-volume chiral continuum limit.
If this were the case, then the introduction of a non-zero fermion mass $m$ would produce bound states with non-zero masses $M$ governed by the mass anomalous dimension $\ga_m^{\star}$ at the IR fixed point~\cite{Miransky:1998dh, Lucini:2009an, DelDebbio:2010ze}.
At leading order, all hadron masses (and decay constants~\cite{DelDebbio:2010jy}) should scale as a power law with the same exponent,
\begin{equation}
  \label{eq:powerLaw}
  \begin{split}
    M_X & = C_X m^{1 / (1 + \ga_m^{\star})} \\
    F_X & = D_X m^{1 / (1 + \ga_m^{\star})}.
  \end{split}
\end{equation}
Lattice calculations further break conformal symmetry through the introduction of a finite volume and finite UV cut-off (inverse lattice spacing)~\cite{Appelquist:2011dp, DelDebbio:2013qta, Cheng:2013xha}, leading to effects that we will not address in this work.

\begin{figure*}[htbp]
  \includegraphics[width=0.45\linewidth]{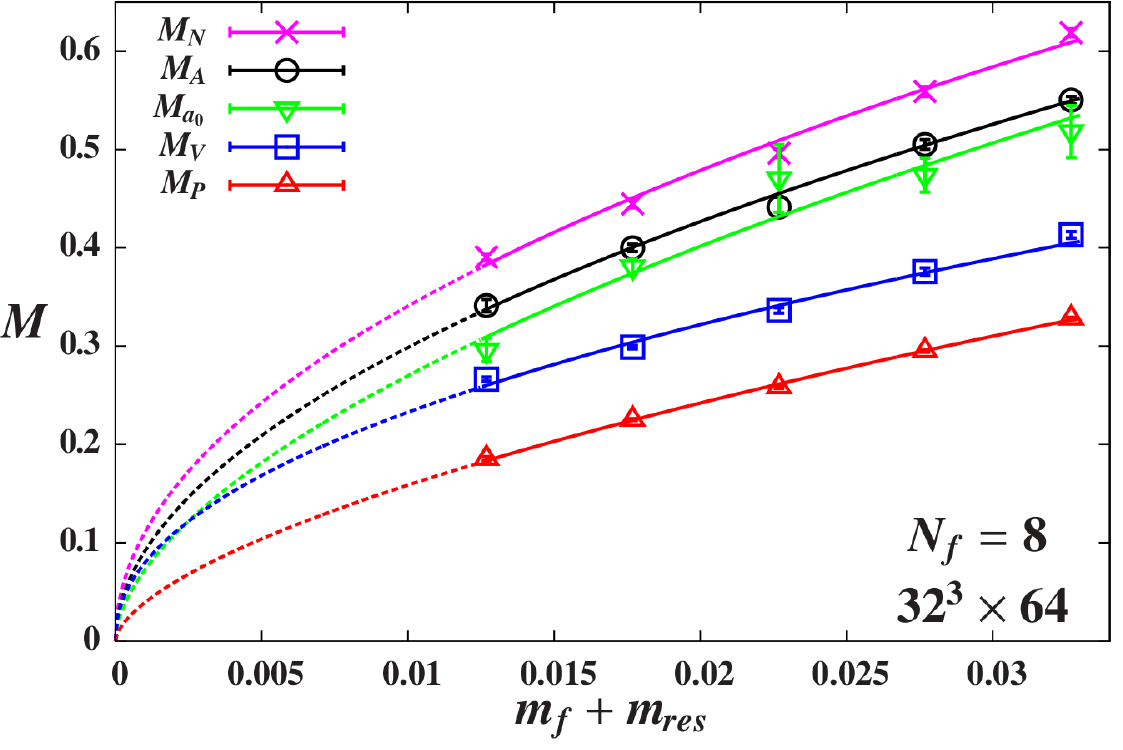}\hfill
  \includegraphics[width=0.45\linewidth]{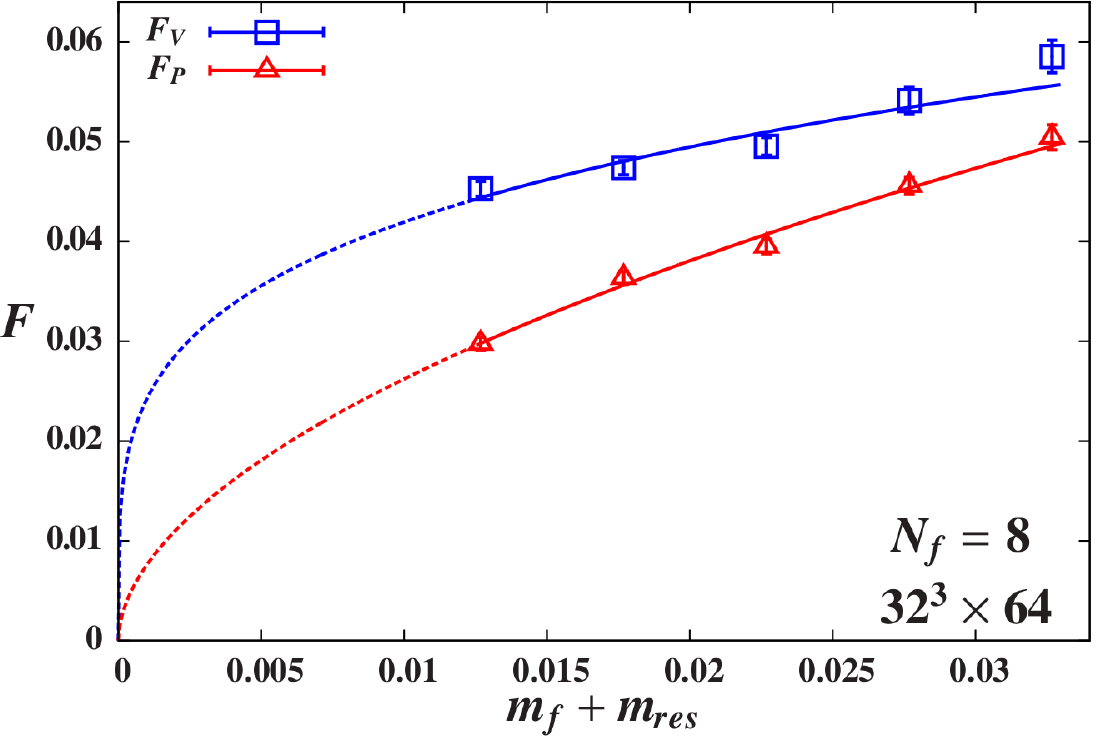}
  \caption{\label{fig:powerFits}Fits of our $32^3\X64$ results for hadron masses (left) and decay constants (right) to the power-law forms in \protect\eq{eq:powerLaw} motivated by mass-deformed IR conformality.  The lines are solid throughout the fit range $0.01 \leq m_f \leq 0.03$.}
\end{figure*}

As in the previous subsection, we consider the simplest possible fits motivated by the IR-conformal hypothesis, to the power-law forms in \eq{eq:powerLaw}.
\fig{fig:powerFits} shows the results, which correspond to the $\ga_m^{\star}$ and \chidof~listed in the second column of \tab{tab:best-gamma}.
Attempts to fit the axial-vector decay constant $F_A$ to a power-law do not succeed, and we omit this observable from the figure and table.
The vector decay constant $F_V$ also leads to unphysically large $\ga_m^{\star} > 2$; the tension between the data and the power-law fit form can be seen in the right panel of \fig{fig:powerFits}.

In \tab{tab:best-gamma} we see that the quality of the other power-law fits is comparable to that of the corresponding linear fits considered in the previous subsection, with a similar range of $1 \lsim \chidof \lsim 10$.
Not surprisingly, the observables with larger \chidof~from power-law fits tend to produce smaller \chidof~from linear fits.
Omitting the lightest point $m_f = 0.01$ from the power-law fits reduces \chidof~for some, but not all, of the observables.

\begin{table*}[htbp]
  \begin{center}
    \begin{tabular}{c|cc|cc|cc|cc}
      \hline
                    & \multicolumn{2}{|c|}{~$0.01 \leq m_f \leq 0.03$~} & \multicolumn{2}{|c|}{~$0.01 \leq m_f \leq 0.025$~}  & \multicolumn{2}{|c|}{~$0.015 \leq m_f \leq 0.03$~}  & \multicolumn{2}{|c}{~$0.015 \leq m_f \leq 0.025$~}  \\
      ~Observable~  & \multicolumn{2}{|c|}{$\text{d.o.f.} = 3$}         & \multicolumn{2}{|c|}{$\text{d.o.f.} = 2$}           & \multicolumn{2}{|c|}{$\text{d.o.f.} = 2$}           & \multicolumn{2}{|c}{$\text{d.o.f.} = 1$}            \\
      \hline
      $M_P$         & 0.64(2)     & 2.5                                 & 0.67(3)     & 2.4                                   & 0.63(2)     & 3.2                                   & 0.66(3)     & 4.4                                   \\
      $M_V$         & 1.14(5)     & 8.7                                 & 1.31(8)     & 6.8                                   & 0.91(6)     & 1.0                                   & 0.99(10)    & 0.9                                   \\
      $M_A$         & 0.94(6)     & 0.7                                 & 0.97(9)     & 1.0                                   & 0.91(7)     & 0.9                                   & 0.92(11)    & 1.7                                   \\
      $M_{a_0}$     & 0.74(16)    & 1.3                                 & 0.67(18)    & 1.7                                   & 1.00(28)    & 0.6                                   & 0.98(36)    & 1.3                                   \\
      $M_N$         & 1.04(4)     & 7.8                                 & 1.22(7)     & 4.2                                   & 0.83(6)     & 3.0                                   & 0.97(11)    & 2.8                                   \\
      $F_P$         & 0.86(10)    & 1.6                                 & 0.92(13)    & 2.1                                   & 0.88(16)    & 2.4                                   & 1.03(24)    & 3.8                                   \\
      $F_V$         & 3.20(49)    & 2.8                                 & 4.07(86)    & 1.7                                   & 2.12(44)    & 1.6                                   & 2.75(86)    & 1.5                                   \\
      \hline
    \end{tabular}
  \end{center}
  \caption{\label{tab:best-gamma}Mass anomalous dimension $\ga_m^{\star}$ and \chidof~from power-law chiral extrapolations of the spectrum data in Tables~\protect\ref{tab:fit_results_32c} and \protect\ref{tab:nucmass}, using the forms in \protect\eq{eq:powerLaw} motivated by mass-deformed IR conformality and considering several different fit ranges.}
\end{table*}

More significantly, the predicted mass anomalous dimension varies over a wide range $0.6 \lsim \ga_m^{\star} \lsim 1.1$ depending on the observable.
While this would suggest that the conformal hypothesis breaks down when confronted with our data, we also see that for most observables $\ga_m^{\star}$ shows significant sensitivity to the fit range.
Taking into account the $\sim\!20\%$ systematic uncertainties suggested by this sensitivity removes much of the tension between different observables, especially given our limited data and the other neglected systematic effects mentioned above.
While these results do not support the hypothesis of mass-deformed infrared conformality, neither do they strongly disfavor it.
The most reliable conclusion we can make is that if the 8-flavor theory were IR conformal, then it would possess a relatively large mass anomalous dimension $0.6 \lsim \ga_m^{\star} \lsim 1.1$.

\begin{figure*}[htbp]
  \includegraphics[width=0.45\linewidth]{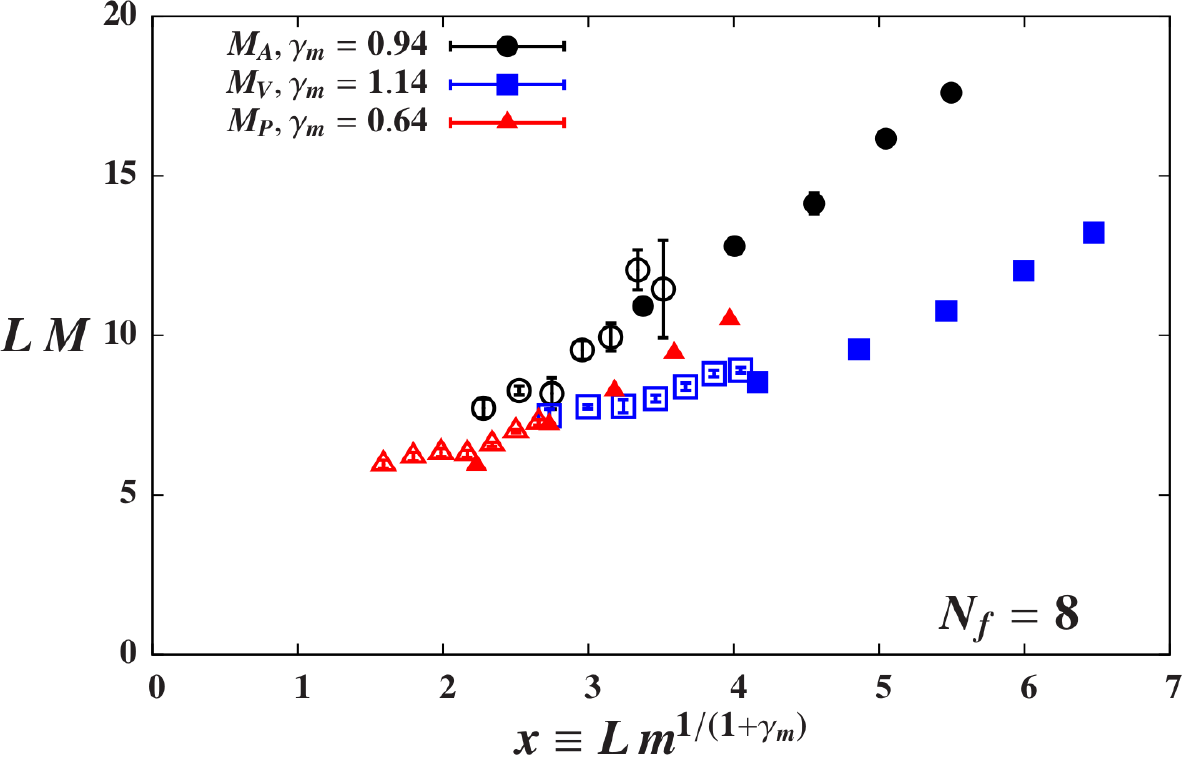}\hfill
  \includegraphics[width=0.45\linewidth]{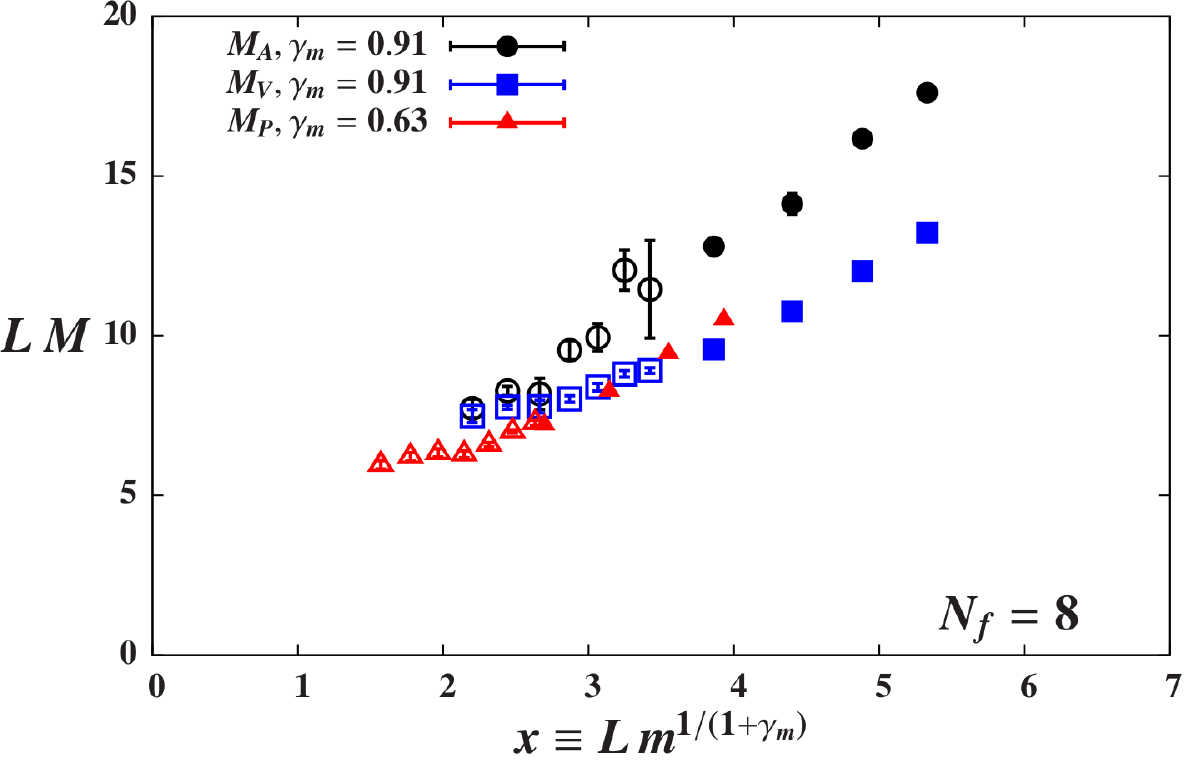}
  \caption{\label{fig:FSS}Testing the consistency of the mass anomalous dimensions $\ga_m$ predicted by power-law fits to $32^3\X64$ data (\protect\tab{tab:best-gamma}) with finite-size scaling.  Filled points are from $32^3\X64$ lattices, while empty points are from $16^3\X32$.  The plot on the left retains all five $32^3\X64$ points with $0.01 \leq m_f \leq 0.03$, while that on the right includes only the four heavier points and uses the corresponding $\ga_m$ from \protect\tab{tab:best-gamma}.}
\end{figure*}

We can attempt to test these power-law fit results by checking their consistency with finite-size scaling.
Considering a mass-deformed IR-conformal theory in a finite spatial volume $L^3$, finite-size scaling states that the hadron masses $M_H$ depend on the scaling variable $x \equiv L m^{1 / (1 + \ga_m)}$, as
\begin{equation}
  M_H = L^{-1} f_H(x).
\end{equation}
As above, we are neglecting several potential complications~\cite{DelDebbio:2013qta, Cheng:2013xha}, so we will not require that our data for different observables all scale with the same fixed $\ga_m^{\star}$.
In \fig{fig:FSS} we plot our $16^3\X32$ and $32^3\X64$ results for $L M_A$, $L M_V$ and $L M_P$ as functions of $x$.
Because there is little or no overlap between the data sets from our two different volumes, we cannot use standard finite-size scaling techniques~\cite{Bhattacharjee:2001am, Houdayer:2004lt} to obtain additional estimates for the mass anomalous dimension.
Instead, we simply use the $\ga_m$ values from \tab{tab:best-gamma} as input, and observe that the $16^3\X32$ points appear to form reasonably continuous extensions of the $32^3\X64$ data that produced these predictions through the power-law fits discussed above.

\section{\label{sec:pbp}Chiral condensate enhancement} 
In composite Higgs models, the chiral condensate \pbp plays the role of the Higgs vacuum expectation value, generating masses for the standard model fermions $f$ through dimension-six interactions of the form $\psibar\psi \fbar f$.
Similar dimension-six couplings $\fbar_i f_i \fbar_j f_j$ generate flavor-changing neutral currents subject to stringent experimental constraints~\cite{Chivukula:2010xz}.
A generic way to satisfy these constraints is for the value of \pbp to be large compared to the symmetry breaking scale $F$.
Such condensate enhancement is conjectured to occur for chirally broken theories near the conformal window~\cite{Holdom:1981rm, Yamawaki:1985zg, Appelquist:1986an}.
In this section we investigate the dimensionless ratio $\pbp / F^3$ for $N_f = 8$, comparing this system with our previous results for $N_f = 2$ and 6~\cite{Appelquist:2009ka, Fleming:2013tra}.

From the leading-order $\chi$PT expression
\begin{equation}
  M_P^2 F_P^2 = 2m\pbp_m
\end{equation}
(the Gell-Mann--Oakes--Renner relation), we can identify three observables that reduce to $\pbp / F^3$ in the chiral limit:
\begin{align}
  X^{(CF)} & = \frac{\pbp_m}{F_P^3} \cr
  X^{(CM)} & = \frac{(M_P^2 / 2m)^{3 / 2}}{\pbp_m^{1 / 2}} \label{eq:pbpF} \\
  X^{(FM)} & = \frac{M_P^2}{2m F_P}. \nonumber
\end{align}
The subscript on $\pbp_m$ indicates that this quantity (like $M_P$ and $F_P$) is evaluated at non-zero fermion mass $m$, in contrast to the chiral-limit values \pbp and $F$.
We discussed $M_P$ and $F_P$ at length in the \secref{sec:spectrum}, and \tab{tab:pbp} presents our results for $\pbp_m$, normalized per flavor (see also Figs.~\ref{fig:pbp} and \ref{fig:therm_16nt32} in the Appendix).
Because we measure $\pbp_m$ directly, these data are dominated by a UV-divergent term $\propto m / a^2$.
As a consequence, the three ratios in \eq{eq:pbpF} have significantly different values in the range of $m$ we can access on $32^3\X64$ lattices.

\begin{table}[htbp]
  \begin{tabular}{ccc}
    \hline
            & \multicolumn{2}{c}{$\pbp_m$}        \\
    $m_f$   & $16^3\X32$      & $32^3\X64$      \\
    \hline
    ~0.010~ &  ---            & ~0.015460(17)~  \\
     0.015  &  ---            &  0.021645(25)   \\
     0.020  & ~0.027023(26)~  &  0.027724(17)   \\
     0.025  &  0.033072(56)   &  0.033742(24)   \\
     0.030  &  0.039124(74)   &  0.039697(11)   \\
     0.035  &  0.045327(62)   &  ---            \\
     0.040  &  0.051396(43)   &  ---            \\
     0.045  &  0.057279(59)   &  ---            \\
     0.050  &  0.063092(66)   &  ---            \\
    \hline
  \end{tabular}
  \caption{\label{tab:pbp}Results for direct measurements of the chiral condensate $\pbp_m$ from $16^3\X32$ and $32^3\X64$ ensembles, normalized per flavor as in Refs.~\protect\cite{Appelquist:2009ka, Fleming:2013tra}.}
\end{table}

\begin{figure*}[htbp]
  \includegraphics[width=0.45\linewidth]{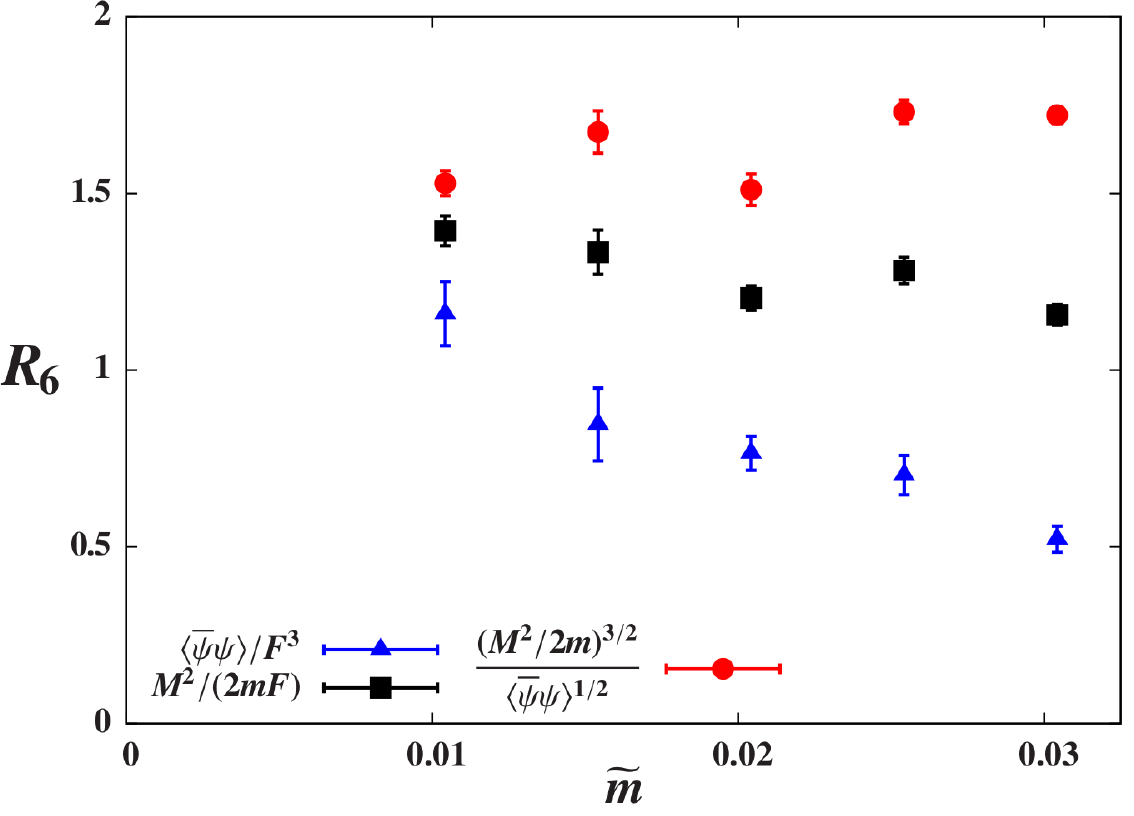}\hfill
  \includegraphics[width=0.45\linewidth]{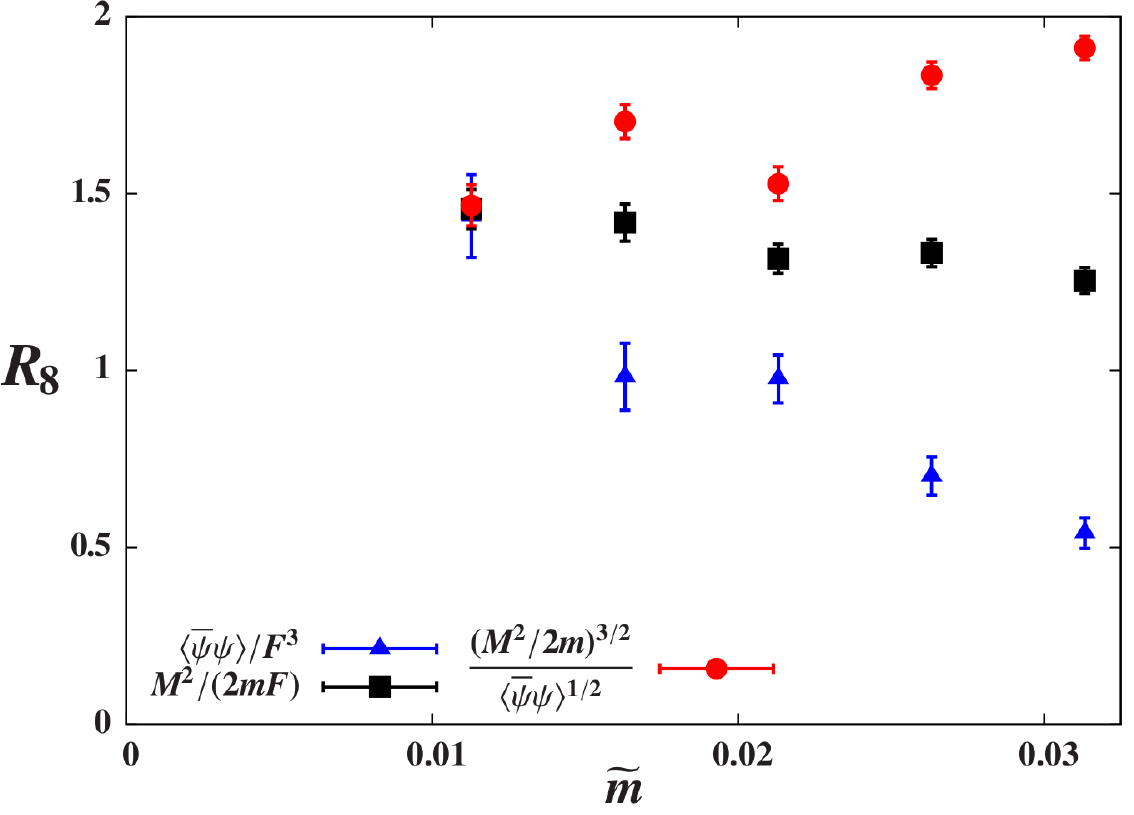}
  \caption{\label{fig:pbpF}Ratios $R_{N_f}^{(IJ)}$ of the three observables $X^{(IJ)}$ in \protect\eq{eq:pbpF} that reduce to $\pbp / F^3$ in the chiral limit, for $N_f = 6$ normalized by $N_f = 2$ (left) and $N_f = 8 / N_f = 2$ (right).  The horizontal axis is the geometric mean $\widetilde m = \sqrt{m_{N_f = 2} m_{N_f = 6, 8}}$.}
\end{figure*}

Following Refs.~\cite{Appelquist:2009ka, Fleming:2013tra} we proceed by comparing each $\pbp / F^3$ observable in \eq{eq:pbpF} with the corresponding 2-flavor quantity, considering the ratios
\begin{equation}
  R_8^{(IJ)} = \frac{X^{(IJ)}(N_f = 8)}{X^{(IJ)}(N_f = 2)},
\end{equation}
where $(IJ)$ enumerates the three constructions $(CF)$, $(CM)$ and $(FM)$.
We take the ratio of 8- and 2-flavor results evaluated with the same input mass $m_f$.
However, these systems have significantly different residual masses (\tab{tab:Nf}), which lead to different $m_{N_f = 2}$ and $m_{N_f = 8}$ in $X^{(CM)}$ and $X^{(FM)}$.
We plot all three ratios $R_8^{(IJ)}$ in the right panel of \fig{fig:pbpF}, using the geometric mean $\widetilde m = \sqrt{m_{N_f = 2} m_{N_f = 8}}$ on the horizontal axis.
The left panel presents our previous results for $R_6^{(IJ)}$ from \refcite{Fleming:2013tra}.

The two plots in \fig{fig:pbpF} are remarkably similar, indicating that the chiral condensate enhancement we observe for $N_f = 8$ is barely larger than what we found for $N_f = 6$.
In particular, the most stable ratio $R^{(FM)}$, which does not involve our direct $\pbp_m$ measurements, increases by only a few percent for $R_8$ compared to $R_6$.
Even so, $\pbp / F^3$ is significantly enhanced for the 8-flavor theory compared to $N_f = 2$.
Based on the 6-flavor joint chiral extrapolation inspired by NLO $\chi$PT in \refcite{Fleming:2013tra}, we obtain the unrenormalized ratio $R_8(\Lambda) \approx 2$ in the $\widetilde m \to 0$ chiral limit, with roughly 10\% statistical uncertainty.

\section{\label{sec:s}Electroweak {\em S}-parameter} 
In this section we consider the electroweak $S$ parameter~\cite{Peskin:1991sw}, following the approach of \refcite{Appelquist:2010xv}.
The $S$ parameter remains one of the most important experimental constraints on electroweak symmetry breaking through new strong dynamics.
$S$ is defined to vanish for the standard model, and its experimental value $S = 0.03(10)$ is consistent with zero~\cite{Baak:2012kk}.
Simply scaling up 2-flavor QCD data to the electroweak scale (and imposing the Higgs mass $M_H = 125$~GeV) would predict $S \approx 0.43$, providing strong evidence that QCD-like technicolor theories are ruled out.

In order to use a many-flavor gauge theory, such as the $N_f = 8$ system under consideration, as the basis of a composite Higgs model, the $S$ parameter must be significantly reduced compared to scaled-up QCD.
In \refcite{Appelquist:2010xv} we observed such a reduction for $N_f = 6$ compared to 2-flavor results, which decreases but does not eliminate the tension with experiment.
Here we repeat this analysis for $N_f = 8$, finding similar results.
We also explore the related issue of parity doubling in the vector and axial-vector channels.

\subsection{Direct analysis of vacuum polarization} 
The $S$ parameter is given by
\begin{equation}
  \label{eq:S}
  S = 4\pi N_D \left.\frac{d}{dQ^2}\Pi_{V - A}(Q^2)\right|_{Q^2 = 0} - \Delta S_{SM},
\end{equation}
where $N_D$ is the number of fermion doublets to which we choose to give chiral electroweak couplings.
Here we fix $N_D = 1$.
$\Pi_{V - A}(Q^2)$ is the transverse component of the difference between vector ($V$) and axial-vector ($A$) vacuum polarization tensors, as a function of euclidean $Q^2 \geq 0$.
Since our domain wall fermion action ensures $Z_V \approx Z_A$, it is straightforward to compute $\Pi_{V - A}(Q^2)$.
These renormalization constants appear since we consider one conserved DWF current and one local current in each correlator~\cite{Schaich:2011qz}.
As first reported by \refcite{Boyle:2009xi}, the use of a single conserved DWF current suffices to ensure that lattice artifacts cancel in the $V$--$A$ difference.
(Such cancellations appear to result from $V$ and $A$ lattice currents forming an exact multiplet under chiral rotations, which is also a feature of the local overlap currents used by \refcite{Shintani:2008qe}.)
Combining one conserved current with a local current reduces computational costs by roughly a factor of $L_s = 16$ compared to using conserved DWF currents exclusively.

To determine $\Pi'(0) \equiv \left.\frac{d}{dQ^2}\Pi(Q^2)\right|_{Q^2 = 0}$, we fit our data for $\Pi_{V - A}(Q^2)$ to a four-parameter Pad\'e-type rational function of the form
\begin{equation}
  \label{eq:pade}
  \Pi(Q^2) = \frac{a_0 + a_1 Q^2}{1 + b_1 Q^2 + b_2 Q^4} = \frac{\sum_{m = 0}^1 a_m Q^{2m}}{1 + \sum_{n = 1}^2 b_n Q^{2n}}.
\end{equation}
The quadratic-in-$Q^2$ denominator in this expression is motivated as a generalization of the single-pole-dominance approximation
\begin{equation}
  \label{eq:singlePole}
  \Pi_{V - A}(Q^2) \approx -F_P^2 + \frac{Q^2F_V^2}{M_V^2 + Q^2} - \frac{Q^2F_A^2}{M_A^2 + Q^2}.
\end{equation}
\eq{eq:pade} is the same fitting function we used in \refcite{Appelquist:2010xv}; subsequent studies~\cite{Aubin:2012me, Golterman:2013vca} have since provided more systematic support for using such rational functions to fit the $Q^2$-dependence of vacuum polarization functions.

Finally, the subtraction of $\Delta S_{SM}$ in \eq{eq:S} removes from the spectrum the three NGBs eaten by the W and Z, and sets $S = 0$ for the standard model with Higgs mass $M_H = 125$~GeV.
Since we have not yet carried out the computationally demanding calculation of the (flavor-singlet scalar) Higgs mass in our lattice studies, we take
\begin{align}
  \Delta S_{SM} = & \frac{1}{12\pi} \int_{4M_P^2}^{\infty} \frac{ds}{s}\left[1 - \left(1 - \frac{M_{V0}^2}{s}\right)^3\Theta(s - M_{V0}^2)\right] \nonumber \\
                  & - \frac{1}{12\pi}\log\left(\frac{M_{V0}^2}{M_H^2}\right).                                                                 \label{eq:DeltaS}
\end{align}
The first term in \eq{eq:DeltaS} would be appropriate if $M_H$ were comparable to the TeV-scale vector meson mass $M_{V0}$; the second term corrects this for the physical $M_H = 125$~GeV~\footnote{Recently \refcite{Foadi:2012bb} argued that a 125-GeV composite Higgs may result from radiative corrections to a state with an intrinsic mass $\sim$600~GeV.  This effect may decrease the second term of \eq{eq:DeltaS}, moving our predictions closer to the experimental value for the $S$ parameter.}.

\begin{figure}[htbp]
  \includegraphics[width=\linewidth]{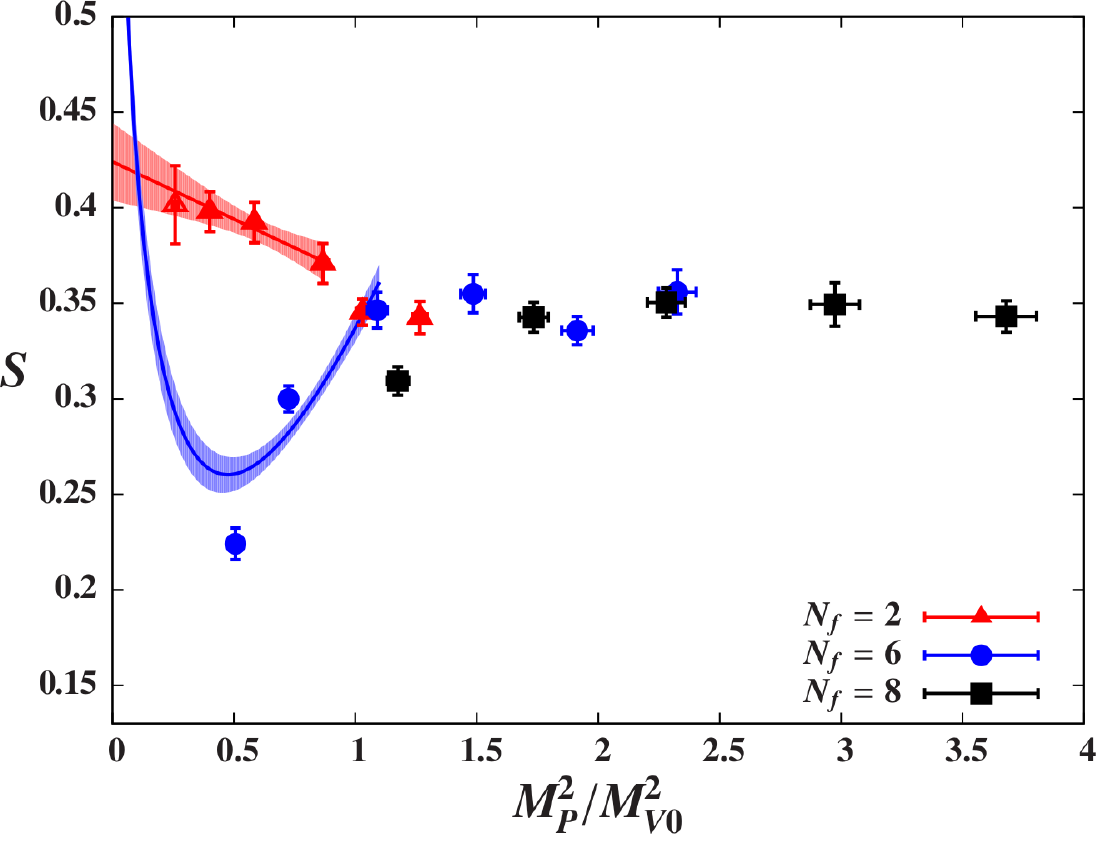}
  \caption{\label{fig:S}Electroweak $S$ parameter with $M_H = 125$~GeV, for $N_f = 2$, 6 and 8 with $N_D = 1$ fermion doublet assigned chiral electroweak couplings in \protect\eq{eq:S}.  Our results for $N_f = 2$ and 6 were previously published in Refs.~\protect\cite{Appelquist:2010xv, Schaich:2011qz}.}
\end{figure}

Computing $S$ for fixed $m$ from Eqs.~\ref{eq:S} and \ref{eq:DeltaS}, employing the thermalization cuts and jackknife blocks listed in \tab{tab:sim_pars}, produces the 8-flavor results shown in \fig{fig:S}.
This figure also includes the $N_f = 2$ and 6 results previously published in \protect\refcite{Appelquist:2010xv}, which we update to use $M_H = 125$~GeV rather than $M_H \sim 1000$~GeV.
As in previous sections, we plot $S$ vs.\ $M_P^2 / M_{V0}^2$ in order to provide a more direct comparison between the three different theories.

The $S$ parameter is only well defined in the chiral limit $M_P^2 / M_{V0}^2 \to 0$.
However, chiral symmetry breaking with $N_f$ light but massive flavors produces $N_f^2 - 1$ PNGBs.
To obtain the phenomenological $S$ parameter, we must consider a chiral limit in which only three of these PNGBs become exactly massless NGBs to be identified with the longitudinal components of the W and Z.
The other $N_f^2 - 4$ PNGBs must remain massive enough to have evaded experimental observation.
(These PNGBs are all pseudoscalars, not to be identified with the 125~GeV Higgs, which comes from the flavor-singlet scalar spectrum that we have not yet investigated.)

For $N_f = 2$ this requirement simply reduces to the linear $M_P^2 / M_{V0}^2 \to 0$ extrapolation shown in \fig{fig:S}, which produces the non-perturbative result $S = 0.42(2)$, in agreement with the scaled-up QCD value $S \approx 0.43$ for $M_H = 125$~GeV.
When $N_f > 2$, keeping all the fermion masses degenerate in the chiral limit would give rise to additional massless NGBs that make a logarithmically divergent contribution to $S$, proportional to $\log\left(M_{V0}^2 / M_P^2\right)$.
The blue band in \fig{fig:S} fits the three $N_f = 6$ data points with the smallest $M_P^2 / M_{V0}^2 \lsim 1$ to the corresponding chiral form~\cite{Schaich:2011qz}.
In a realistic context, the $N_f^2 - 4$ PNGBs remain massive, due to standard model and other interactions, which break this degeneracy.

For $N_f = 8$, we cannot access $M_P^2 / M_{V0}^2 < 1$ on $32^3\X64$ lattice volumes, making this sort of chiral fit unreasonable.
Even so, in \fig{fig:S} we can observe the beginning of a similar reduction in our 8-flavor results for $S$.
The Edinburgh-style plot in \fig{fig:Edin} suggests that these results should be safe from finite-volume distortions.
(The lightest $N_f = 2$ and $N_f = 6$ points in \fig{fig:S} use $m_f = 0.005$ and are omitted from \fig{fig:Edin}; finite-volume effects may be significant for this 6-flavor point.)
Because $N_f = 8$ is closer to the conformal window, we would expect this reduction to end up more significant than that for $N_f = 6$ at smaller $M_P^2 / M_{V0}^2$, but this cannot be determined from our current lattice results.

\subsection{\label{sec:parity}Vector and axial-vector parity doubling} 
The expected decrease in the $S$ parameter for systems near the conformal window is related to the onset of parity doubling between the vector and axial-vector channels.
This can be seen in \eq{eq:singlePole}, which follows from the dispersion relation
\begin{equation}
  \Pi_{V - A}(Q^2) = \frac{Q^2}{12\pi} \int_0^{\infty} \frac{ds}{\pi} \left[\frac{R_V(s) - R_A(s)}{s + Q^2}\right] - F_P^2,
\end{equation}
upon approximating each spectral function $R(s)$ by a single pole,
\begin{equation}
  \label{eq:poles}
  \begin{split}
    R_V(s) & \approx 12\pi^2 F_V^2\delta(s - M_V^2) \\
    R_A(s) & \approx 12\pi^2 F_A^2\delta(s - M_A^2).
  \end{split}
\end{equation}
Parity doubling in this context amounts to the statement that $R_V(s) \approx R_A(s)$, so that $\Pi_{V - A}'(0) \approx 0$.

\begin{figure*}[htbp]
  \includegraphics[width=0.45\linewidth]{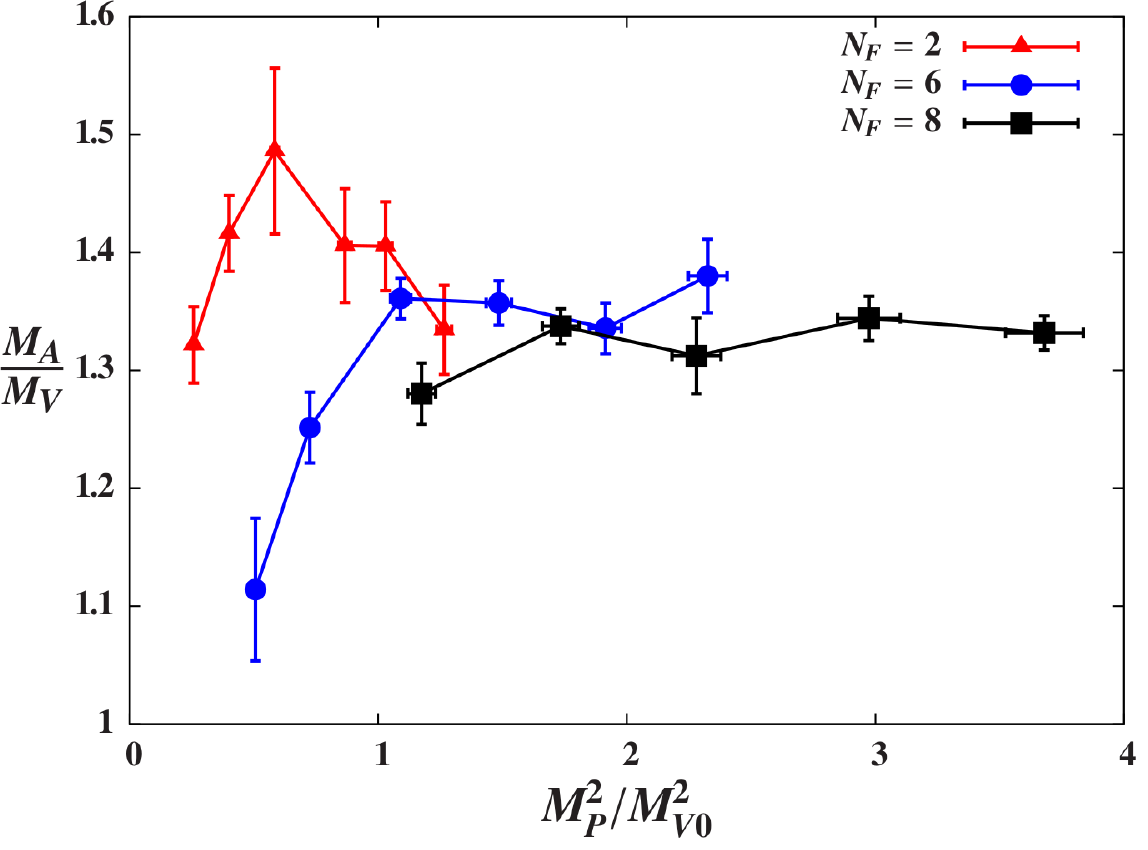}\hfill
  \includegraphics[width=0.45\linewidth]{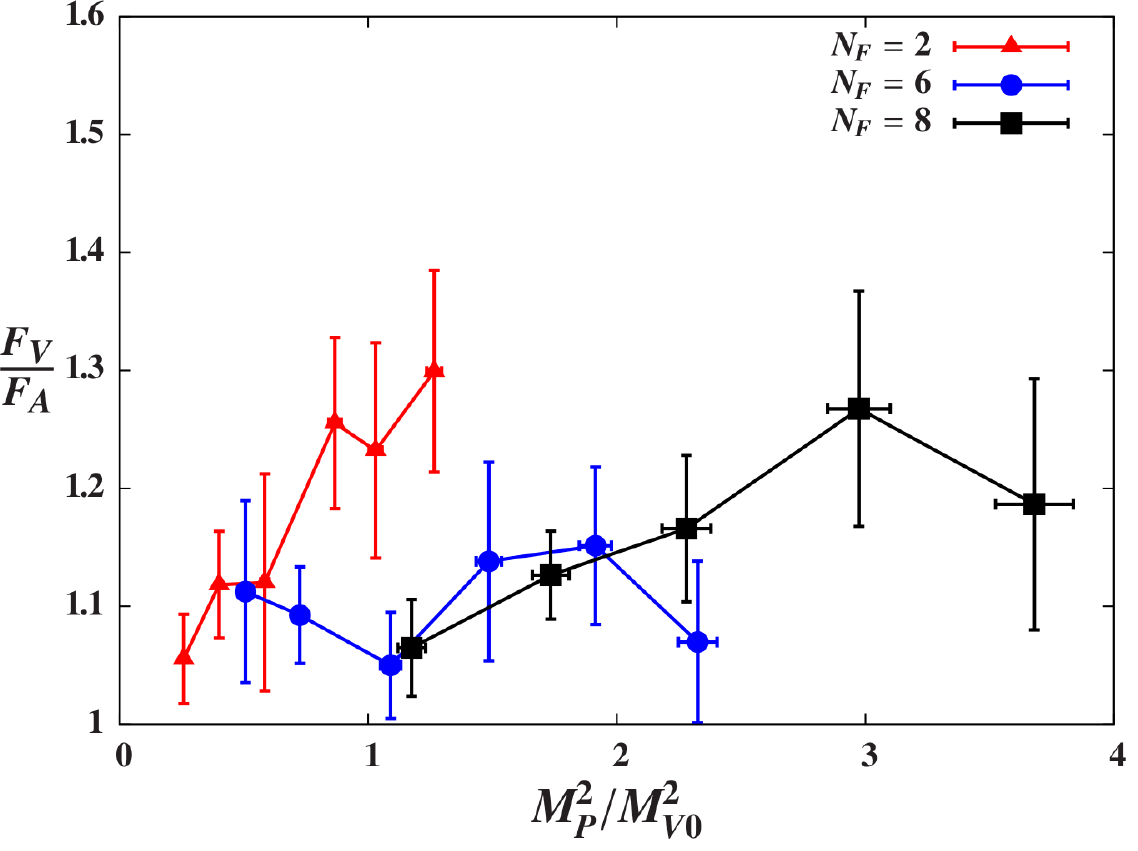}
  \caption{\label{fig:parDoub}Parity doubling in the vector and axial-vector channels: $M_A / M_V$ (left) and $F_V / F_A$ (right) plotted against $M_P^2 / M_{V0}^2$ for $N_f = 2$, 6 and 8.  Lines connect points to guide the eye.  Our results for $N_f = 2$ and 6 are from \protect\refcite{Appelquist:2010xv}.}
\end{figure*}

Instead of attempting to reconstruct the full spectral functions $R_V(s)$ and $R_A(s)$, we focus on the lightest vector and axial-vector states discussed in \secref{sec:spectrum}.
From the single-pole approximation in \eq{eq:poles}, we see that in addition to roughly degenerate masses $M_V \approx M_A$, parity doubling also requires approximately equal decay constants $F_V \approx F_A$.
In \fig{fig:parDoub} we plot the ratios $M_A / M_V$ and $F_V / F_A$ for $N_f = 2$, 6 and 8, again as functions of $M_P^2 / M_{V0}^2$.
Comparing these plots with \fig{fig:S}, we see that reductions in our direct measurements of the electroweak $S$ parameter are indeed associated with increased parity doubling.
Both the 6- and 8-flavor ratios tend to be smaller than the $N_f = 2$ results at comparable $M_P^2 / M_{V0}^2$, with their smallest values corresponding to the points at which $S$ decreases.

\section{\label{sec:conclusion}Conclusion} 
We have presented results from lattice investigations of SU(3) gauge theory with $N_f = 8$ degenerate domain wall fermions in the fundamental representation.
With ensembles of gauge configurations covering a range of fermion masses on two lattice volumes $32^3\X64$ and $16^3\X32$ at a single gauge coupling $\beta = 1.95$, we studied the light hadron spectrum, chiral condensate \pbp and electroweak $S$ parameter.
We tuned $\beta$ to approximately match IR scales with those from our previous studies of $N_f = 2$ and 6.
The resulting coupling $\beta = 1.95$ leads to a relatively large residual mass $\mres = 0.002684(7)$ with $L_s = 16$.

At this gauge coupling, our $32^3\X64$ lattices appear to be large enough to keep finite-volume effects under control for input fermion masses $m_f \geq 0.01$, as illustrated by the Edinburgh-style plot in \fig{fig:Edin}.
The ratio $M_V / M_P$ steadily increases as we approach the chiral limit, suggesting spontaneous chiral symmetry breaking.
Simple linear chiral extrapolations of our spectrum results, motivated by the hypothesis of spontaneous chiral symmetry breaking, tend to be of reasonable quality.
The main exception is the pseudoscalar mass squared, for which a large \chidof~and significantly non-zero chiral-limit value indicate tension with leading-order $\chi$PT.
Considering the possibility that the 8-flavor theory is IR conformal, we find that power-law chiral extrapolations motivated by the hypothesis of mass-deformed IR conformality are of comparable quality.
These power-law fits predict a relatively large mass anomalous dimension $0.6 \lsim \ga_m^{\star} \lsim 1.1$.

To explore chiral condensate enhancement, we studied three observables that reduce to $\pbp / F^3$ in the chiral limit.
Compared to $N_f = 2$, our 8-flavor results indicate a significant enhancement of $\pbp / F^3$, with an unrenormalized $N_f = 8 / N_f = 2$ ratio $R_8(\Lambda) \approx 2$ in the chiral limit.
Our results for the 8-flavor condensate enhancement are remarkably similar to our previous findings for $N_f = 6$~\cite{Fleming:2013tra}.
For the $N_f = 8$ electroweak $S$ parameter we also observe behavior similar to the 6-flavor case, with a reduction in $S$ setting in at the smallest masses we can reliably access on $32^3\X64$ lattices.
This reduction in the $S$ parameter is associated with increased parity doubling between the lightest vector and axial-vector states.

Given the recent discovery of a light Higgs particle, the most pressing task for future lattice studies of candidate walking technicolor theories is to investigate the flavor-singlet scalar spectrum.
It remains an open question whether strongly-coupled gauge theories produce a scalar state with the properties of the 125~GeV Higgs.
Although such studies are difficult and computationally expensive, these efforts are now underway for the 8-flavor SU(3) theory, with promising initial results reported by the LatKMI Collaboration~\cite{Aoki:2014oha}.
The LatKMI investigations using staggered fermions motivate complementary studies with domain wall fermions, but the larger cost of DWF make these calculations even more challenging.
The 8-flavor DWF ensembles we have generated do not appear to possess sufficient statistics to permit robust glueball analyses, but we are currently exploring fermion-line-disconnected DWF calculations, to judge whether such flavor-singlet scalar studies are practical.

\section*{Acknowledgements} 
We thank M.~Buchoff, A.~Hasenfratz and T.~DeGrand for useful discussions, and A.~Rago for assistance with glueball analyses.
We thank the Lawrence Livermore National Laboratory (LLNL) Multiprogrammatic and Institutional Computing program for Grand Challenge allocations and time on the LLNL BlueGene/L supercomputer, along with funding from LDRD~10-ERD-033.
Additional numerical analyses were carried out on clusters at LLNL.
M.F.L.\ was partially supported by SciDAC-3 and Argonne Leadership Computing Facility at Argonne National Laboratory under contract DE-AC02-06CH11357, and by the Brookhaven National Laboratory Program Development under grant PD13-003.
D.S.\ was supported by the U.S.~Department of Energy (DOE) under Grant Nos.~DE-SC0008669 and DE-SC0009998.
R.C.B., C.R.\ and E.W.\ were supported by DOE grant DE-SC0010025.
In addition, R.C.B., C.R.\ and O.W.\ acknowledge the support of NSF grant OCI-0749300.
G.T.F.\ and G.V.\ were supported by NSF grant PHY11-00905.
E.R., C.S.\ and P.V.\ acknowledge the support of the DOE under Contract DE-AC52-07NA27344 (LLNL).

\begin{figure*}[ht]
  \includegraphics[width=0.45\linewidth]{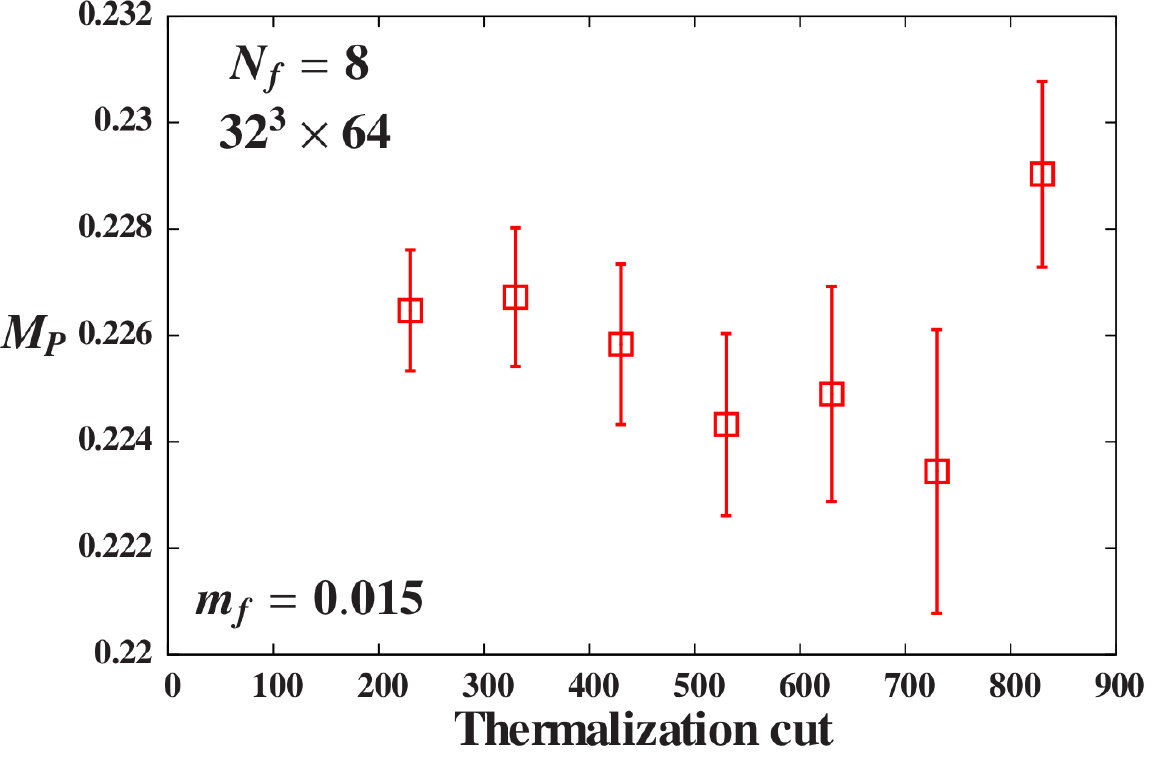}\hfill
  \includegraphics[width=0.45\linewidth]{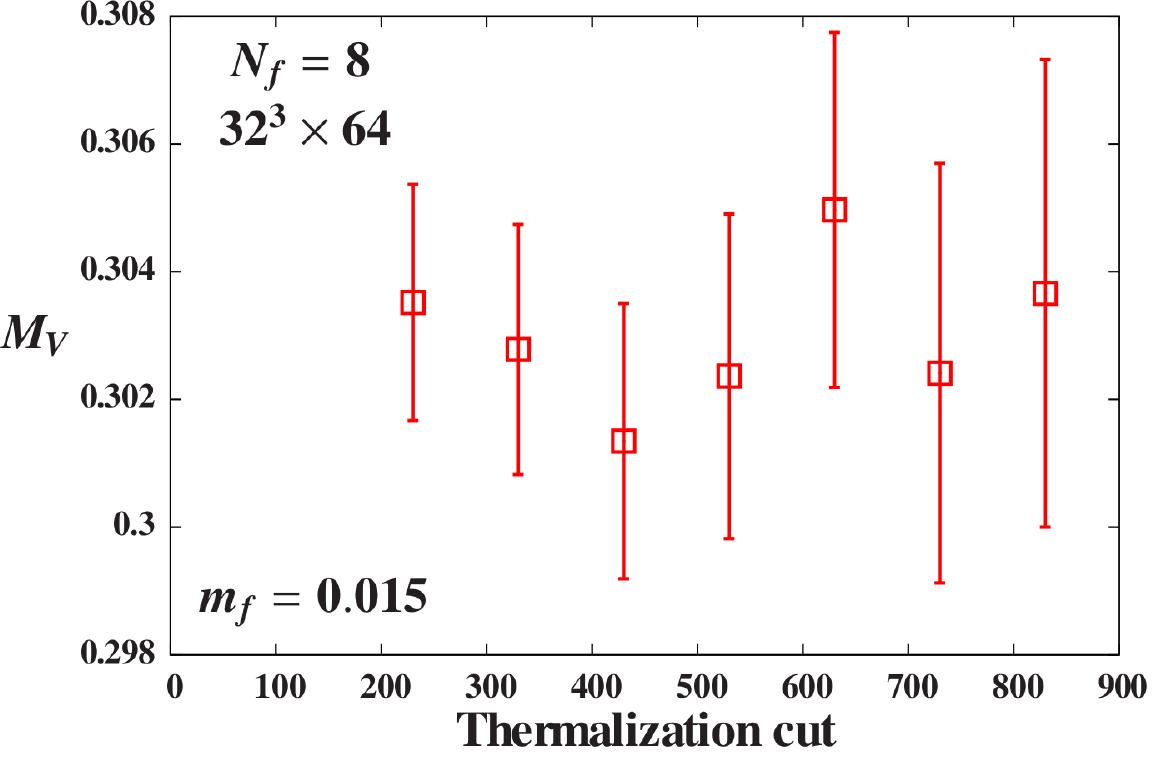}
  \caption{\label{fig:Mtherm}$M_P$ (left) and $M_V$ (right) as functions of the thermalization cut for the disordered-start $32^3\X64$ $m_f = 0.015$ ensemble.  The masses reach plateaus for a thermalization cut around 500 MD trajectories.}
\end{figure*}

\appendix{\subsection*{Appendix: Thermalization, auto-correlations and topological charge evolution}} 
In this appendix we provide additional information about the thermalization and auto-correlations of our lattice ensembles, which motivate our use of 50-trajectory jackknife blocks and the thermalization cuts listed in \tab{tab:sim_pars}.
We set the thermalization cuts in \tab{tab:sim_pars} by monitoring spectral results as functions of the thermalization cut.
Representative results for $M_P$ and $M_V$ on the disordered-start $32^3\X64$ $m_f = 0.015$ ensemble are shown in \fig{fig:Mtherm}.
As the thermalization cut increases the spectral quantities initially evolve, but their central values reach plateaus once the system has thermalized.
(The error bars continue to grow as the amount of remaining data decreases, and we attribute the jump in the final $M_P$ point to the limited data; in these plots we use jackknife blocks of 20 MD trajectories.)
For this ensemble the results reach plateaus once the thermalization cut is around 500 MD trajectories.

\begin{figure*}[ht]
  \includegraphics[width=0.45\linewidth]{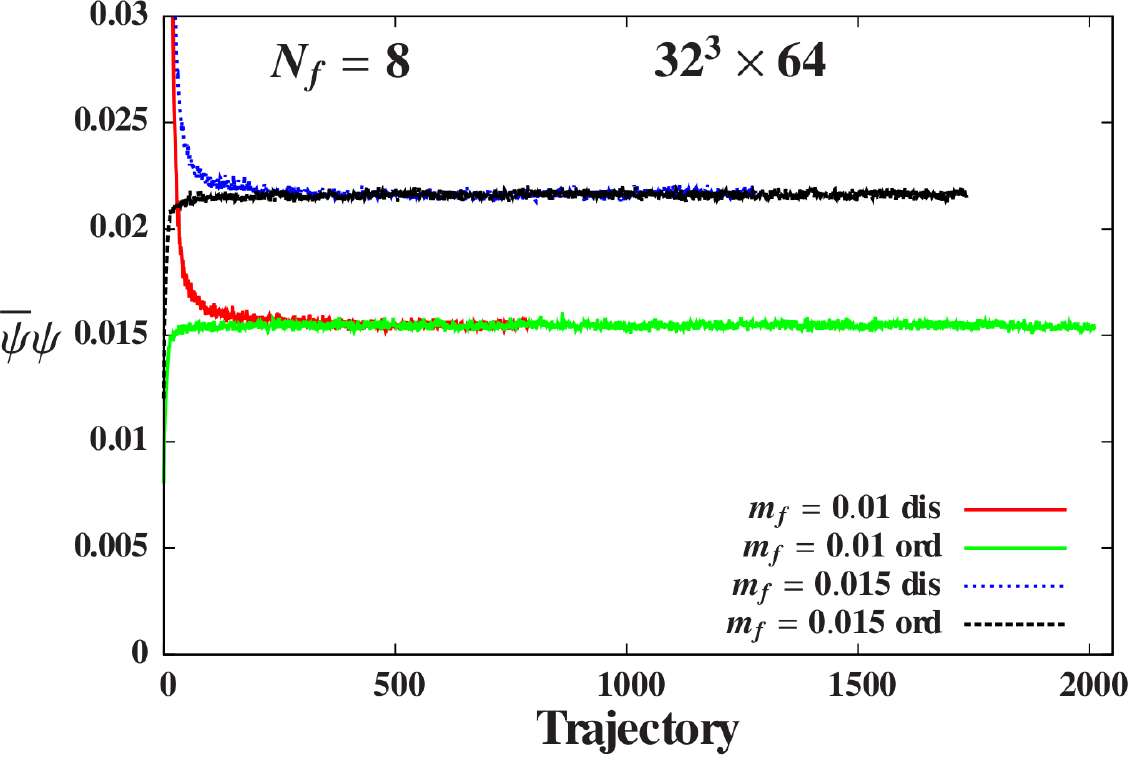}\hfill
  \includegraphics[width=0.45\linewidth]{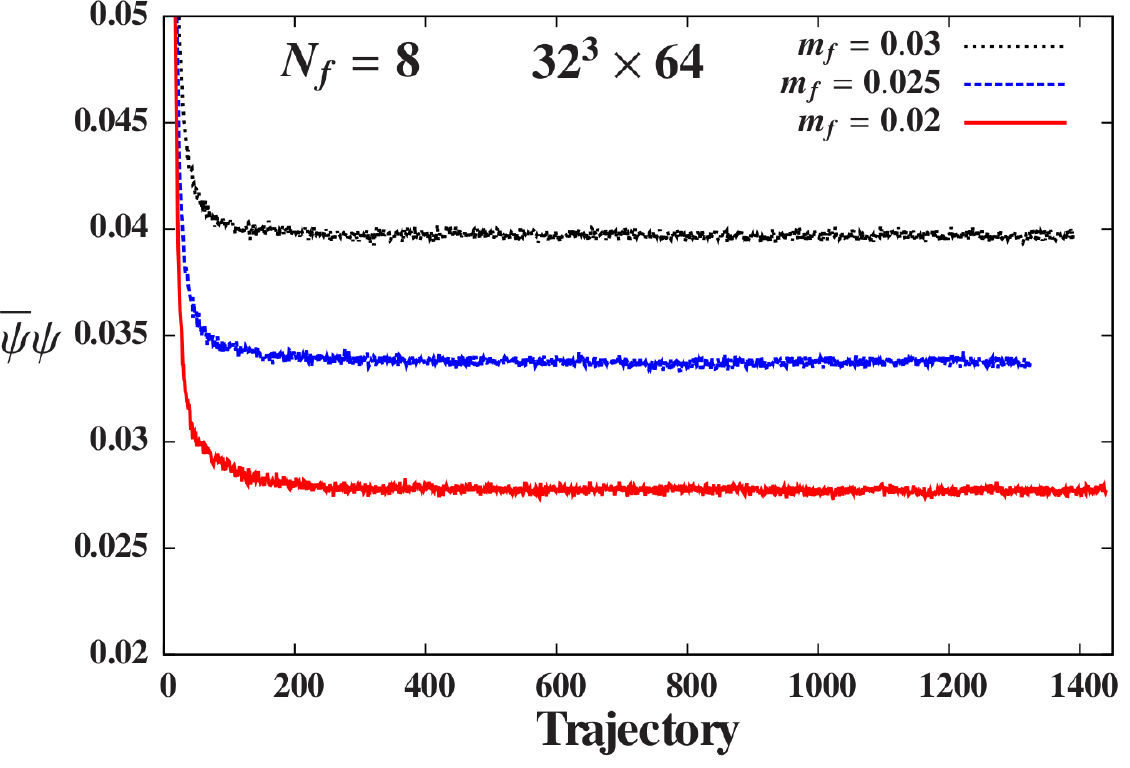}
  \caption{\label{fig:pbp}Time-series plots of the chiral condensate $\psibar\psi$ for $32^3\X64$ ensembles.  Left: $m_f = 0.01$ and 0.015 including both the ordered and disordered starts.  Right: $m_f = 0.02$, 0.025 and 0.03.}
\end{figure*}

\begin{figure*}[ht]
  \includegraphics[width=0.45\linewidth]{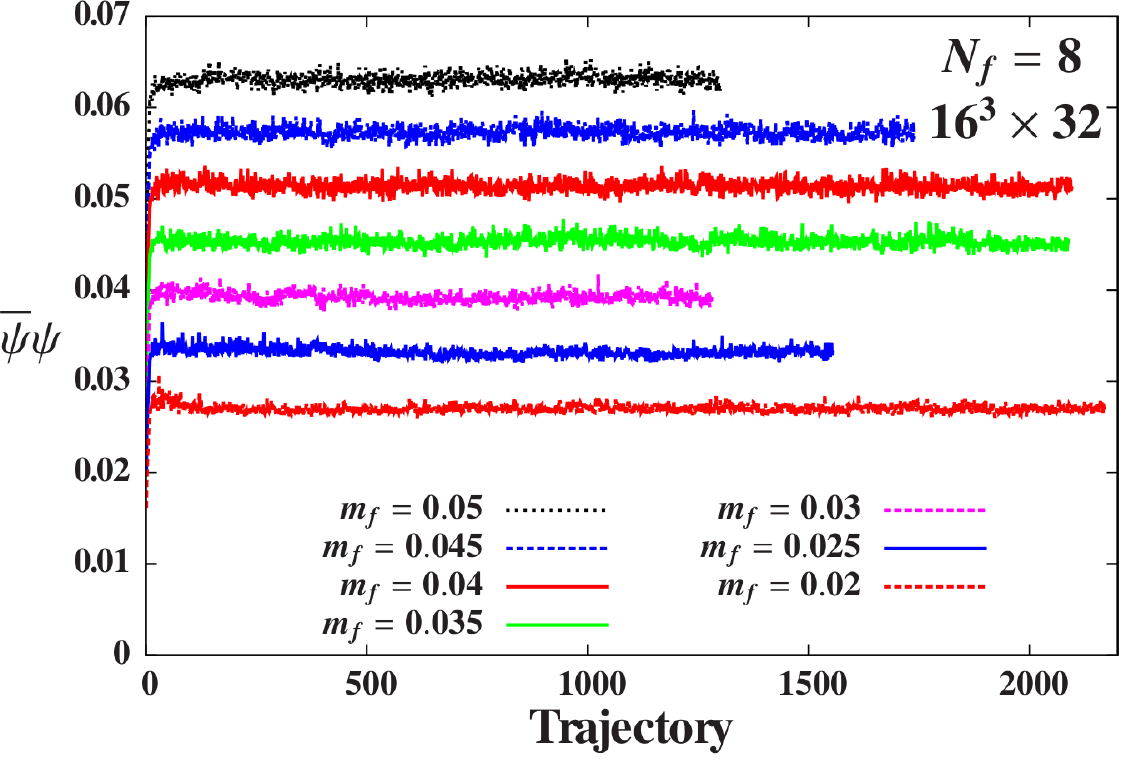}\hfill
  \includegraphics[width=0.45\linewidth]{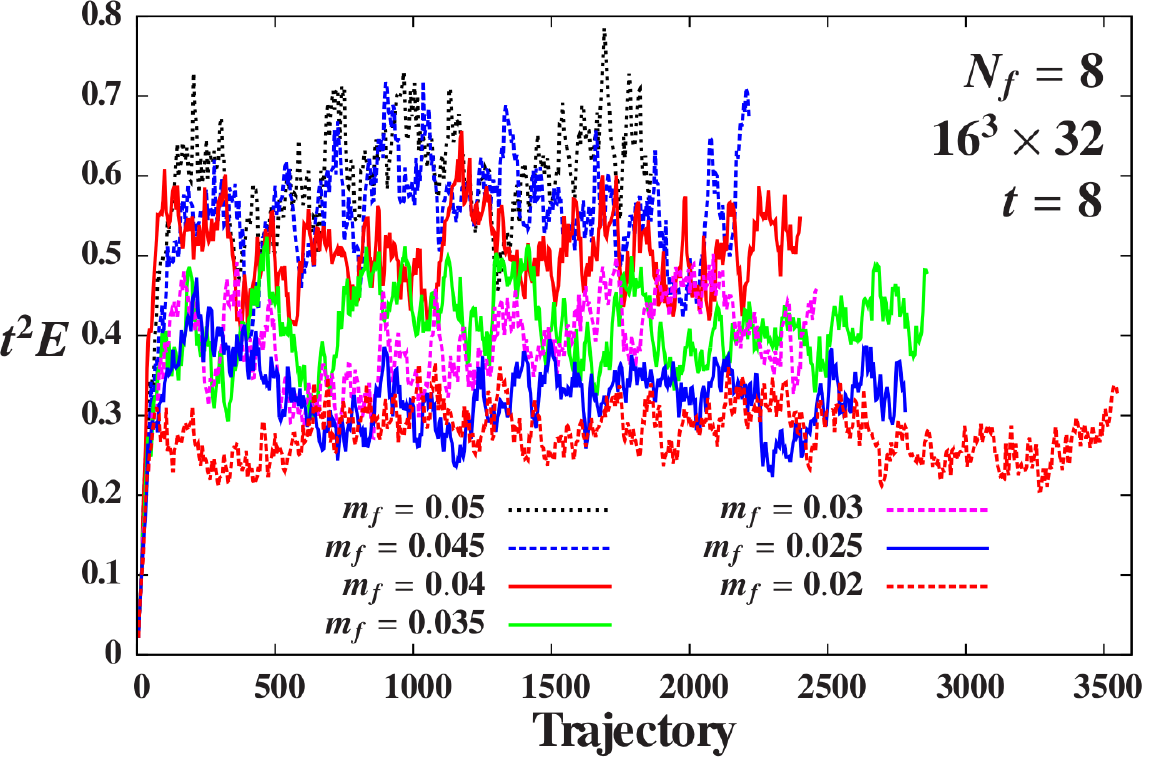}
  \caption{\label{fig:therm_16nt32}Time-series plots for $16^3\X32$ ensembles.  Left: The chiral condensate $\psibar\psi$.  Right: $t^2 E$ from \protect\eq{eq:Wflow} after running the Wilson flow to $\sqrt{8t} = L / 2$.}
\end{figure*}

A complementary way to estimate thermalization is to inspect time-series plots of appropriate observables.
The chiral condensate $\psibar\psi$ is often used for this purpose.
In Figs.~\ref{fig:pbp} and \ref{fig:therm_16nt32} for our $32^3\X64$ and $16^3\X32$ ensembles, respectively, we see that $\psibar\psi$ appears to thermalize quite rapidly, within tens of MD trajectories for $L = 16$, and a couple hundred MD trajectories for $L = 32$.
Subsequent fluctuations in $\psibar\psi$ are too small to be readily visible in these figures.

\begin{figure*}[ht]
  \includegraphics[width=0.45\linewidth]{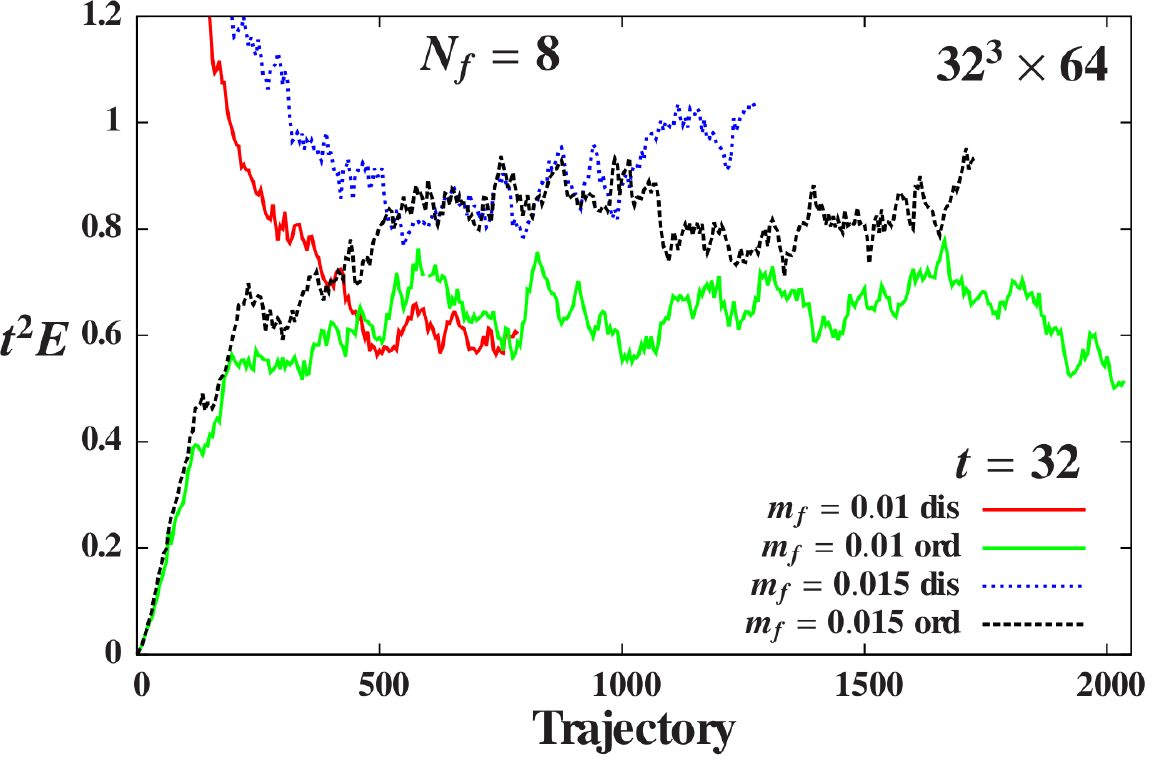}\hfill
  \includegraphics[width=0.45\linewidth]{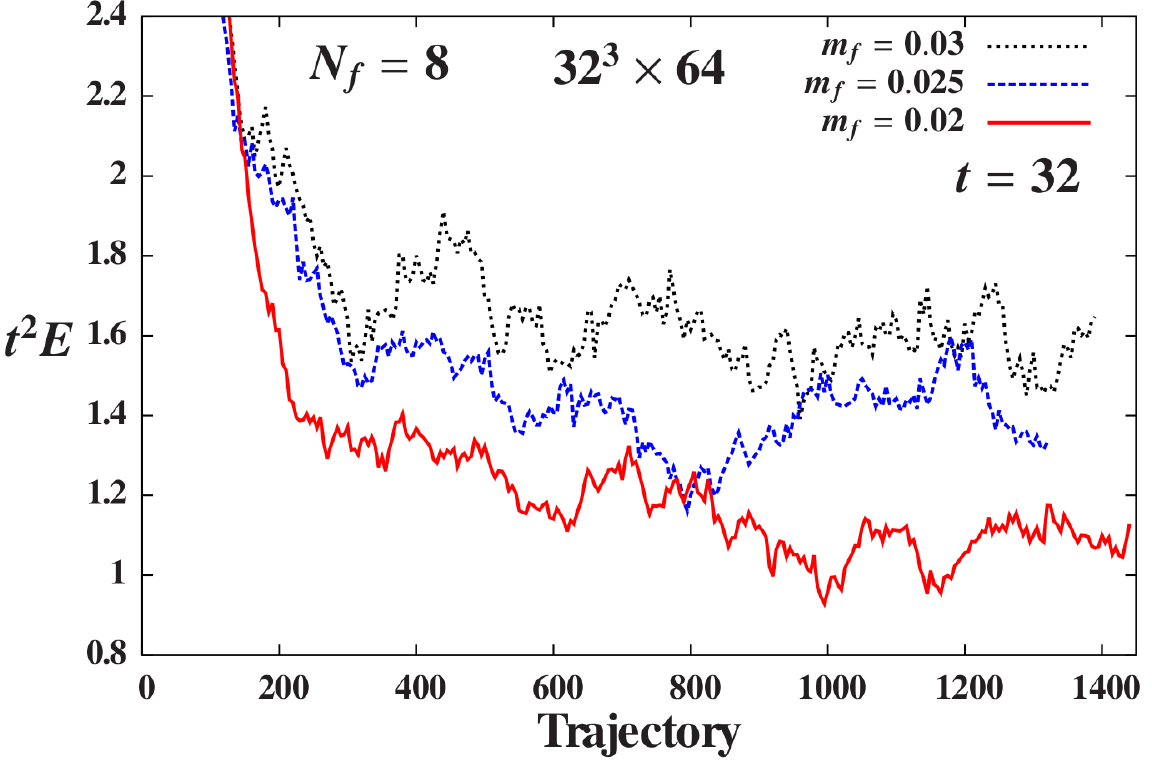}
  \caption{\label{fig:therm}Time-series plots of $t^2 E$ for $32^3\X64$ ensembles, from \protect\eq{eq:Wflow} after running the Wilson flow to $\sqrt{8t} = L / 2$.  Left: $m_f = 0.01$ and 0.015 including both the ordered and disordered starts.  Right: $m_f = 0.02$, 0.025 and 0.03.}
\end{figure*}

An alternative observable, which we have found to be more sensitive to thermalization than is $\psibar\psi$~\cite{Schaich:2013eba}, is the quantity $t^2 E(t)$ measured after long Wilson flow times $t$.
The Wilson flow is a continuous and reversible smoothing operation, integrated to cover a radius $\sim \sqrt{8t}$ (cf.~\cite{Luscher:2013vga} for a recent review).
This smoothing provides long-distance quantities that are not too noisy, and are relatively inexpensive to compute.
After running the Wilson flow for some flow time $t$, the energy $E$ and topological charge $Q$ are simply
\begin{equation}
  \label{eq:Wflow}
  \begin{split}
    E & = -\frac{1}{2}\mbox{ReTr}\left[F_{\mu\nu}F^{\mu\nu}\right] \\
    Q & = \frac{1}{32\pi^2}\mbox{ReTr}\left[\epsilon_{\mu\nu\rho\sigma} F^{\mu\nu}F^{\rho\sigma}\right],
  \end{split}
\end{equation}
both of which we determine from the clover-leaf definition of $F_{\mu\nu}$.
Figs.~\ref{fig:therm} and \ref{fig:therm_16nt32} present time-series plots for $t^2 E(t)$ for our $32^3\X64$ and $16^3\X32$ ensembles, respectively, which show more gradual thermalization than do those for $\psibar\psi$.
The thermalization cuts we would estimate from this Wilson flow observable are generally consistent with those in \tab{tab:sim_pars} that we set by monitoring spectral quantities.


\begin{figure*}[ht]
  \includegraphics[width=0.45\linewidth]{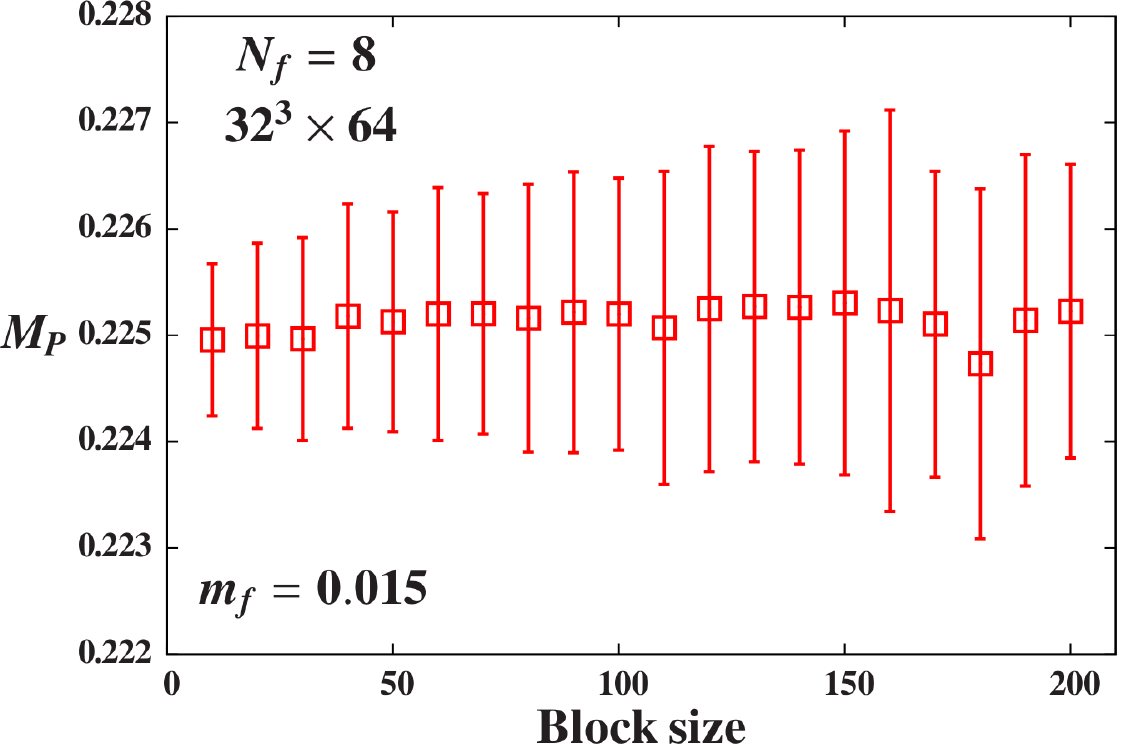}\hfill
  \includegraphics[width=0.45\linewidth]{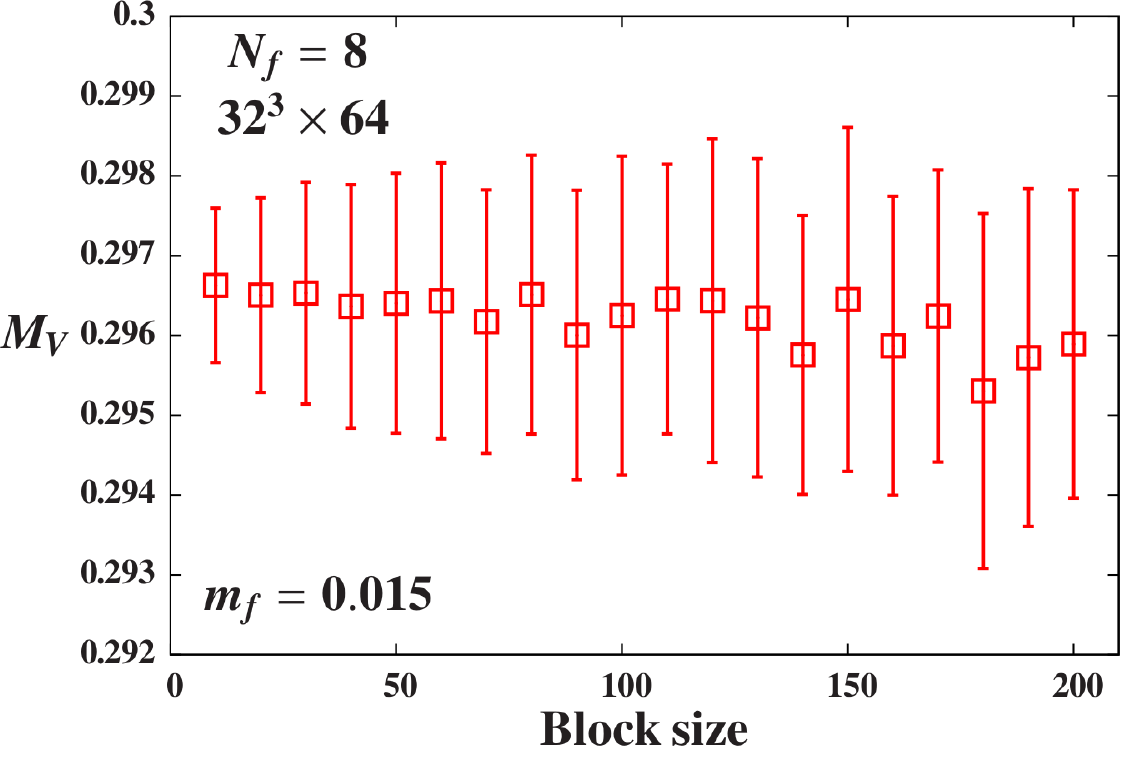}
  \caption{\label{fig:Mblock}$M_P$ (left) and $M_V$ (right) as functions of the jackknife block size for the ordered-start $32^3\X64$ $m_f = 0.015$ ensemble.  The size of the error bars appears to stabilize around block sizes of roughly 100 MD trajectories.}
\end{figure*}

The fluctuations of $t^2 E(t)$ following thermalization are also more significant than the fluctuations of $\psibar\psi$, which could allow us to estimate auto-correlation times.
That said, we choose the size of our jackknife blocks from the common approach of measuring masses as functions of the block size.
\fig{fig:Mblock} presents representative results for $M_P$ and $M_V$ on the ordered-start $32^3\X64$ $m_f = 0.015$ ensemble.
As the block size increases, auto-correlations are removed and so the statistical uncertainties increase towards their true values.
For this ensemble the size of the error bars appears to stabilize around block sizes of roughly 100 MD trajectories.
Due to the limited amount of data available for some of our other $32^3\X64$ ensembles, we are not able to use jackknife blocks larger than 50 MD trajectories.
\fig{fig:Mblock} indicates the extent to which our statistical uncertainties may be underestimated as a consequence.
In general, 100-trajectory jackknife blocks produce uncertainties roughly 25\% larger than those from the 50-trajectory blocks we use, which led us to increase our error estimates by this factor.

\begin{figure*}[ht]
  \includegraphics[width=0.45\linewidth]{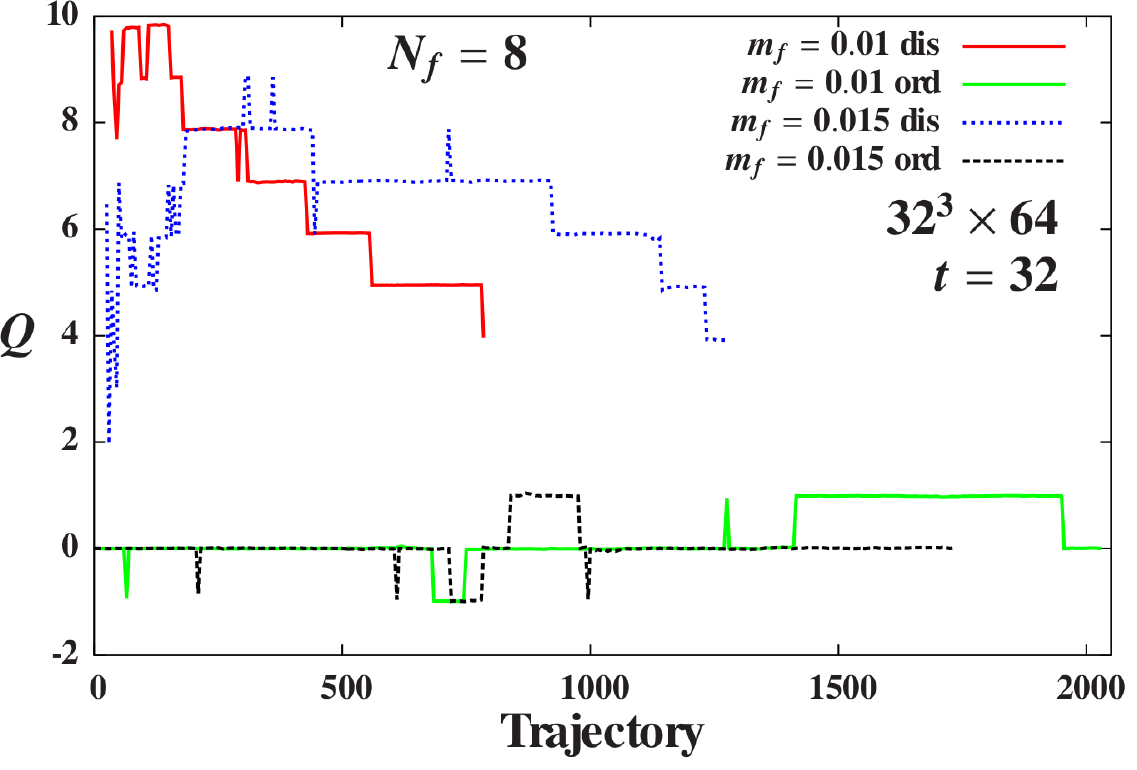}\hfill
  \includegraphics[width=0.45\linewidth]{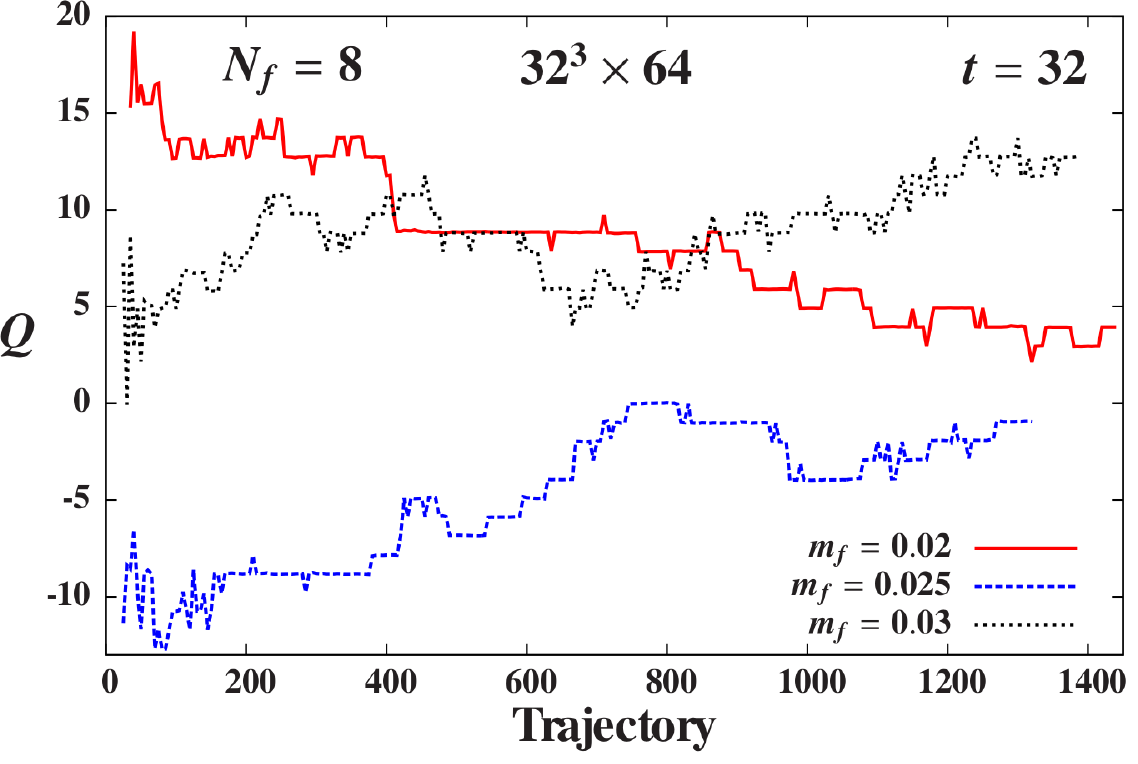}
  \caption{\label{fig:topo}Time-series plots of the topological charge for $32^3\X64$ ensembles, from \protect\eq{eq:Wflow} after running the Wilson flow to $\sqrt{8t} = L / 2$.  Left: $m_f = 0.01$ and 0.015 including both the ordered and disordered starts.  Right: $m_f = 0.02$, 0.025 and 0.03.}
\end{figure*}

The global topological charge $Q$ is well known to exhibit particularly severe auto-correlations, especially as the lattice spacing decreases~\cite{Alles:1996vn, DelDebbio:2004xh} or $N_f$ increases~\cite{Appelquist:2009ka, Appelquist:2012nz}.
Conveniently, the Wilson flow measurements of $t^2 E(t)$ discussed above also determine the topological charge, as shown in \eq{eq:Wflow}.
By considering long flow times $t$, we obtain nearly-integer values for $Q$ on our $32^3\X64$ lattices, which we plot in \fig{fig:topo}.
Clearly, none of our ensembles exhibit the gaussian topological charge distribution around $Q = 0$ that we desire.
However, in contrast to the 10-flavor case where the topological charge is almost entirely frozen~\cite{Appelquist:2012nz}, we observe frequent tunneling, especially for larger $m_f$.
In the future, it would be interesting to determine topological susceptibilities and estimate auto-correlation times from these topological charge time series, following the maximum-likelihood approach proposed by \refcite{Brower:2014bqa}.

\raggedright
\bibliographystyle{apsrev}
\bibliography{refs}
\end{document}